\documentstyle[prd,aps,eqsecnum,preprint,tighten]{revtex}
\begin{document}
\draft
\title
{Exact solutions in the Yang-Mills-Wong theory\\}
\author{B. P. Kosyakov 
\cite{byline}\\}
\address{Theoretical Division, Russian Federal Nuclear Center, 
Sarov 607190, Russia\\}
\maketitle
\begin{abstract}
This paper discusses in a systematical way exact retarded solutions to 
the classical SU$(N)$ Yang-Mills equations with the source composed of several 
colored point particles.
A new method of finding such solutions is reviewed. 
Relying on features of the solutions, a toy model of quark binding is 
suggested.
According to this model, quarks forming a hadron are influenced by no 
confining force in spite of the presence of a linearly rising term of 
the potential.
The large-$N$ dynamics of quarks conforms well with Witten's phenomenology.
On the semiclassical level, hadrons are color neutral in the Gauss law sense.
Nevertheless, a specific multiplet structure is observable in the 
form of the Regge sequences related to infinite-dimensional unitary 
representations of SL$(4,R)$ which is shown to be the color gauge group of the 
background field generated by any hadron.
The simultaneous consideration of SU$(N)$, SO$(N)$, and Sp$(N)$ as gauge groups
offers a plausible explanation of the fact that clusters containing two or 
three quarks are more stable than multiquark clusters.
\end{abstract}

\widetext
\section{Introduction}

In this paper, we look into the Yang-Mills-Wong (YMW) theory, a classical 
non-Abelian gauge model describing closed systems of several spinless colored 
point particles interacting with a gluon field.
Emphasis is put on features of exact solutions, some of which were obtained 
previously \cite{kosy,kos},  nd others are discussed for the first time. 

A solvable nonlinear model is of interest by itself.
Further still, one may expect that exact solutions of the YMW theory will be 
useful in studying nonperturbative vacua of quantum chromodynamics (QCD). 
We get, at the least, a toy model of bound states in QCD.

It may appear at first glance that classical particles have nothing 
to do with real quarks, but closer inspection casts doubt on this belief.
Indeed, a medium of the normal nuclear density offers a fertile ground for 
creations and annihilations of quark pairs.
Nevertheless, the quark-antiquark sea is largely suppressed in hadrons. 
According to Zweig's rule, a quark and an antiquark with opposite quantum
numbers defy their annihilation.
Such persistence of particles is typical for the classical picture.

The classicality of constituent quarks is difficult to understand, but   
one can describe it explicitly.
The bulk of the hadron phenomenology is grasped by planar diagrams, which
implies in particular that world lines of valence quarks are subjected to 
neither bifurcation nor termination in the Feynman path integral, unless 
hadrons collide or decay. 
Thus, to a good approximation, the number of bound quarks remains fixed.

Starting from the QCD Lagrangian, Wong was able to show \cite{wong} that
the behavior of a quark in the limit $\hbar\to 0$ is governed by the 
classical equation of motion of a spinless colored point particle.
\footnote{The present model is not quite that obtained in the original Wong 
approach.
We deal with arbitrary number of color particles while Wong's procedure 
is matched with a single particle; the situations are apparently different if 
one keeps in mind the nonlinear field dynamics.}
Alternative methods of ``dequantization" \cite{frad} lead to classical 
actions of the spinning particle with Wong's action as a constituent.
Thus the classical limit of the quark dynamics remains to be completed.
Wong's dynamics may provisionally be regarded as the simplest reasonable 
approximation to the limiting theory.
In addition, it would be appropriate to use spinless particles as the starting 
point of the bound quark description, for measurements of the polarized 
proton structure in deep inelastic leptoproduction indicate that quarks 
carry only a small fraction of the spin of the nucleon \cite{emc}.
The Wong particles will hereafter be called quarks though this name is rather 
conventional and should not be confused with the standard QCD term, 
taking into account that the Wong color charges are in the adjoint while quark 
fields are in the fundamental representation of the gauge group.

As is well known, the classical limit of QCD is related to the limit of 
large number of colors \cite{ya}.
Substituting SU$(3)_c$ by SU$(N)$ and going to the limit 
\[N\to\infty,\quad g^2N={\rm const},\] 
't Hooft established that the planar diagrams are dominating in 
this limit \cite{hooft}.
Witten found \cite{wit} that the real hadronic world is qualitatively 
displayed even in the zeroth approximation of the $1/N$ expansion. 
In the limit $N\to\infty$, the vacuum expectation value of the product of 
gauge invariant operators obeys the so-called factorization relation, and 
quantum fluctuations disappear \cite{wit}. 
Thus QCD becomes a classical theory as $N\to\infty$.
We suggest that the large-$N$ YMW theory is intimately related to the 
classical limit of QCD.

Note, however, that the confinement problem is out of the question now. 
Indeed, it is conceivable that quarks constituting a hadron experience an 
attractive constant force originating from a term of the potential $A_\mu$ 
which linearly rises with distance between the quarks, and such a behavior of 
$A_\mu$ is to provide the area law for the Wilson loop functional \cite{wils}. 
Are we correct in interpreting the area law as the evidence of the constant 
attractive force?
As will be shown, an exact classical solution $A_\mu$ with the linearly 
rising term actually exists.
Although this term contributes to the field strength, it produces no force.
The general reason for such a surprising result is the conformal invariance.
The linearly rising term violates the scale invariance.
While such a violation being allowable for the gauge quantities $A_\mu$ and
$F_{\mu\nu}$, it cannot be tolerated for observables.
One may expect a dimensional parameter, measu\-ring a gap in the energy 
spectrum and violating the scale symmetry, to emerge only upon quantization 
leading to anomalies.
Meanwhile exact classical solutions are crucial in learning the symmetry 
of the vacuum. 

One believes two phases of the strong interacting matter to exist, hot 
and cold, which must be distinguished by their symmetry. 
At high temperatures, the asymptotical freedom dominates, hence the 
conventional SU$(3)_c$ symmetry is inherent in the hot phase.
On the other hand, Ne'eman and {\v S}ija{\v c}ki \cite{ne} developed 
an exhaustive phenomenological classification of hadrons on the basis of
infinite-dimensional unitary representations of SL$(4,R)$, which hints that 
SL$(4,R)$ is the cold phase symmetry. 
Where does this SL$(4,R)$ come from?

Coleman \cite{col} argued that the symmetry of the vacuum is the symmetry of 
world.
Given the vacuum invariant under SL$(4,R)$, excitations about it possess the 
same symmetry. 
Since the symmetry of the gluon vacuum is nothing but the symmetry of the 
background field, the responsibility for SL$(4,R)$ rests with the background 
described by a certain solution of the QCD equations in the classical limit.
It is the background generated by quarks in hadrons that provides the 
SL$(4,R)$ relief for gluon excitations.

We will find two classes of exact retarded solutions to the classical 
Yang-Mills (YM) equations.
Solutions of the first class, invariant under SU$(N)$, appear to be related
to the background in the hot phase. 
Solutions of the second class might be treated as the back\-ground generated 
by bound quarks in the cold phase. 
These solutions are complex valued with respect to the Lie algebra su$(N)$, 
but one can convert them to the real form to yield the invariance under
SL$(N,R)$ or its subgroups. 
In particular, the background generated by any three-quark cluster is 
invariant under SL$(4,R)$, and that generated by any two-quark cluster 
is invariant under SL$(3,R)$.

Notice that SL$(4,R)$ of Ne'eman and {\v S}ija{\v c}ki operates in spacetime 
while the present SL$(4,R)$ acts in the color space.
However, we attempt to interweave two arenas by reference to that color 
degrees of freedom may be convertible into spin degrees of freedom, the fact 
discovered by Jackiw and Rebbi, and Hasenfratz and 't Hooft \cite{jare}.

The paper is organized as follows.
Section II outlines the general formalism of the YMW theory.
The next section is devoted to a justification of the Ansatz whereby we seek 
exact retarded solutions of the YM equations with the source composed of 
several arbitrarily moving quarks.
Finding such solutions is traced by the simplest example of the 
single-quark source, Sec. IV.
Properties of the background generated by two-quark sources are reviewed 
in Sec. V. 
Fields generated by arbitrary number of color particles, and their 
features are considered in Sec. VI. 
We show that bound quarks are affected by no external force as $N\to\infty$.
The equation of motion of a dressed quark is discussed, and its exact 
solution in the absence of external forces is given.
The large-$N$ dynamics of quarks is shown to conform with Witten's 
phenomenology.
Moreover, a step forward can be made by the simultaneous consideration of 
SU$(N)$, SO$(N)$, and Sp$(N)$ as gauge groups, which offers a plausible 
explanation of the fact that clusters composed of two or three quarks are 
more stable than multiquark clusters.
The stability of the solutions is the subject of Sec. VII.
We conclude that free quarks are ruled out by a consistency reasoning.
Issues concerning the semiclassical quantization and the resulting picture 
are considered in Sec.  VIII.
The fulfilment of the Wilson criterion has been confirmed.
We show that any hadron is color-neutral in the sense of the Gauss law.
Nevertheless, a certain multiplet structure is observable.
These multiplets are described by infinite-dimensional unitary representations 
of SL$(4,R)$, the gauge group of the background field generated by any hadron.

\section{General formalism}
We work in Minkowski space with the metric $\eta_{\mu\nu}={\rm diag}
(+,\,-,\,-,\,-)$.
Let us consider classical point particles interacting with the SU$(N)$ 
Yang-Mills field.
The particles will be called quarks and labeled by index $I,\,I =1,\ldots,K$. 
Each quark is assigned a color charge $Q^a_I$ (transforming as the adjoint 
representation of SU$(N)$, the color index $a$ runs from $1$ to $N^2-1$), and 
a bare mass $m^I_0$. 
Any other specification is omitted, so that quarks and antiquarks are 
indistinguishable in the present context. 
Let every quark be moving along a timelike world line $z_I^\mu(\tau_I)$ 
parametrized by the proper time $\tau_I$. 
This gives rise to the current
\begin{equation}
j_\mu (x) =\sum^K_{I=1}\int\!d\tau_I\, Q_I(\tau_I)\,
v^I_\mu (\tau_I)\,\delta^4\bigl [x-z_I (\tau_I)\bigr],
\label{1}\end{equation}                                         
where $Q_I=Q_I^a\,T_a$,\, $T_a$ are generators of SU$(N)$, 
$v^I_\mu\equiv\dot z^I_\mu\equiv dz^I_\mu/d\tau_I$ is the four-velocity of $I$th
quark.
The action is written \cite{bbs} as 
\begin{equation}
S=-\sum_{I=1}^K\,\int d\tau_I\,(m^I_0\,\sqrt{v_\mu^I\,v_I^\mu}+
{\rm tr}\,Z_I\lambda^{-1}_I{\dot\lambda}_I)
-\int d^4x\,{\rm tr}\,\bigl(j_\mu\,A^\mu+{1\over 16\pi}\,
F_{\mu\nu}\,F^{\mu\nu}\bigr).
\label{2}\end{equation}                                          
Here, $\lambda_I=\lambda_I(\tau_I)$ are time-dependent elements of SU$(N)$,
$Z_I=e_I^aT_a$,\, $e_I^a$ being some constants whereby the color charge is
specified, $Q_I=\lambda_I Z_I\lambda_I^{-1}$.
The field strength is 
\[F_{\mu\nu}=\partial_\mu A_\nu-\partial_\nu A_\mu-ig\,[A_\mu,A_\nu],\] 
with $g$ being the coupling constant. 
The middle two terms of Eq. (\ref{2}) can be combined into 
$-\sum{\rm tr}\,Z_I\lambda^{-1}_I D_\tau{\lambda}_I$ with the covariant 
derivative $D_\tau\equiv d/d\tau_I+v_I^\mu A_\mu(z_I)$.
Since $\lambda_I$ responds to a local gauge transformation by 
$\lambda_I\to\lambda'_I=\Omega^{-1}\lambda_I$, the gauge invariance of the 
action is quite clear.
Elements $T_a$ of the Lie algebra su$(N)$ satisfy the commutation relations
\begin{equation}
[T_a,\,T_b] =if_{abc}T^c
\label{3}\end{equation}                                           
with the structure constants $f_{abc}$ of SU$(N)$, and the 
orthonormalization condition
\begin{equation}
{\rm tr}\,(T_a\, T_b) =\delta_{ab}.
\label{4}\end{equation}                                         

Note that the invariance of the action under SU$(N)$ automatically 
entails the invariance under SL$(N,C)$ unless a constraint is 
imposed so as to preserve the real-valueness of the gauge field variables. 
If we have no prior knowledge of the symmetry, it can be 
identified by the structure constants which are present in the action. 
The specific values of $f_{abc}$ entering into Eq. (\ref{3}) 
imply that $S$ is invariant under SU$(N)$. 
However, for any simple complex Lie algebra, there exists a basis, referred to 
as the Cartan basis, such that the structure constants are found to 
be real, antisymmetric and identical to the structure constants of the 
real compact form of this Lie algebra\cite{barr}. 
The basis of su$(N)$ is simultaneously the Cartan basis of its 
complexification sl$(N,C)$. 
Thus the presence of the structure constants of SU$(N)$ in Eq. (\ref{3}) 
needs not be the evidence for that the symmetry of $S$ is 
SU$(N)$; allowing for the complex-valued field variables, we 
enlarge the symmetry up to SL$(N,C)$.

The Euler-Lagrange equations for the action (\ref{2}) are the Yang-Mills 
equations
\begin{equation}
{\cal E}_\mu\equiv D^\nu F_{\mu\nu} +4\pi j_\mu=0
\label{5}\end{equation}                                           
with $D^\nu=\partial^\nu-ig[A^\nu,\,\,\,]$,
the equation of motion of $I$th bare quark
\begin{equation}
\varepsilon_I^\lambda\equiv m_0^I\,a^\lambda_I-
v_\mu^I\,{\rm tr}\,\bigl[Q_I\, F^{\lambda\mu}(z_I)\bigr]
=0,
\label{6}\end{equation}                                           
where $a^\lambda_I\equiv \dot v^\lambda_I$ is the four-acceleration of this 
quark, and the Wong equation
\begin{equation}
\dot Q_I=-ig\,[Q_I,\,v^I_\mu\,A^\mu (z_I)]
\label{8}\end{equation}                                           
describing the evolution of the color charge of $I$th quark. 

It follows from Eq. (\ref{8}) that
\[\frac{d}{d\tau_I}\,{\rm tr}\,Q_I^2= 
-2ig\,{\rm tr}\bigl(Q_I\,[Q_I,\,v^I_\mu A^\mu]\bigr) =0,\]
i.e., the magnitude of $Q_I$ remains unchanged, specifically, $Q_I$ may be 
constant.

The total color charge of a $K$-quark system is defined by 
\begin{equation}
Q=\int_{\Sigma} d\sigma^\mu j_\mu,
\label{10}\end{equation}                                          
where the integral is taken over an arbitrary spacelike hypersurface 
${\Sigma}$, and the domain of integration covers all $K$ points of 
intersection of ${\Sigma}$ with the world lines. 
However, it would be more convenient to do with somewhat narrow class of 
hypersurfaces with a rigidly fixed mutual arrangement of any hypersurface 
and the world lines in the vicinities of their intersections.
Consider, for example, the intersection of a hyperplane by a timelike 
curve at a right angle. 
Such an arrangement can be achieved for any hypersurface $\Sigma$ by replacing 
a small fragment of $\Sigma$ in the vicinity of every intersection point 
$z^I_\mu$ by a fragment of a hyperplane orthogonal to the world line 
at $z^I_\mu$ and smoothing off this piecewise hypersurface. 
The resulting hypersurface will be called {\it locally adjusted} and denoted 
by $\tilde\Sigma$.

Since $j_\mu$ is not a conserved current, the total color charge $Q$ is in 
general hypersurface dependent.
But $Q$ ceases to depend on  $\tilde\Sigma$ if the color charge of each quark 
is constant,
\begin{equation}
\dot Q_I=0,
\label{12}\end{equation}                                          
which imposes certain restrictions on the form of $A_\mu$. 
We will discuss just this case. 

In view of Eq. (\ref{5}), the definition of $Q$ can also be rewritten in terms 
of the field variables:
\begin{equation}
Q={1\over 4\pi}\int_{\tilde\Sigma} d\sigma_\nu D_\mu 
F^{\mu\nu}.
\label{13}\end{equation}                                          

Under the local gauge transformations 
\[A_\mu\to\Omega^{-1}A_\mu\Omega-{i\over g}\,\Omega^{-1}
\partial_\mu\Omega,\]
the covariant derivatives of the field strength transform as
\[D_\mu F^{\mu\nu}\to\Omega^{-1}D_\mu F^{\mu\nu}\Omega,\]
so that Eq. (\ref{5}) is covariant providing $j_\mu$ transforms as
\[j_\mu\to\Omega^{-1}j_\mu\Omega.\]
One could always find such unitary matrix $\Omega$ as to 
diagonalize Hermitean matrix $j_\mu$. 
Since the Lie algebra su$(N)$ is of rank $N -1$, there exist 
$N -1$ diagonal elements $H_i$. 
Thus, without loss of generality, one can set
\begin{equation}
Q_I(\tau_I)=\sum_{i=1}^{N-1} e^i_I(\tau_I)\,H_i.
\label{15}\end{equation}                                          
We will find $e^i_I$ to be constants fixed exactly by the solution itself.

Picking $Q_I$ in the form (\ref{15}) reduces the gauge freedom of $A_\mu$.
The color charges $Q_I$ may thereon be rotated within the Cartan 
subgroup, in particular, through discrete angles associated with permutations 
of $H_i$.
We will see that the diagonalization of $Q_I$ leads to
\begin{equation}
[Q_I,\,v^I_\mu\,A^\mu ]=0
\label{9}\end{equation}                                          
which can be treated as a gauge fixing condition.

The symmetric energy-momentum tensor is 
\[T_{\mu\nu}=\Theta_{\mu\nu}+t_{\mu\nu},\] 
where
\begin{equation}
\Theta_{\mu\nu} = {1\over 4\pi} \,{\rm tr}\,(F_{\mu\alpha}\,
F^\alpha_{{\hskip1.5mm}\nu} +
{1\over 4}\,\eta_{\mu\nu}\,F_{\alpha\beta}\,F^{\alpha\beta}),                   
\label{16}\end{equation}                                          
\begin{equation}
t_{\mu\nu} (x) =\sum^K_{I=1} m_0^I\int d\tau_I\,v^I_\mu 
(\tau_I)\,v^I_\nu(\tau_I)\,\delta^4\bigl[x-z_I (\tau_I)\bigr].  
\label{17}\end{equation}                                          
One can readily verify the Noether identity
\begin{equation}
\partial_\mu T^{\lambda\mu}={1\over 4\pi}\,{\rm tr}\,
({\cal E}_\mu\,F^{\lambda\mu})+\sum_{I=1}^K\int\! d\tau_I\,
\varepsilon_I^\lambda\,\delta^4\bigl[x-z_I(\tau_I)\bigr],
\label{18}\end{equation}                                          
where ${\cal E}_\mu$ and $\varepsilon_I^\lambda$ are the 
LHS's of Eqs. (\ref{5}) and (\ref{6}), respectively. 

As is well known (see , e.g., \cite{alf}), the conformal 
invariance is ensured by the condition
\[T^\mu_{\hskip1.5mm\mu}=0.\]
Leaving aside sophistications regarding the modification of the 
energy-momentum tensor on the quantum level \cite{ccj}, we merely remark 
that the YM sector is conformally invariant for
\[\Theta^\mu_{\hskip1.5mm\mu}=0.\]
Equation (\ref{16}) shows that this condition is met for $D=4$.
Thus, in the 4D case, the spacetime symmetry of the YM 
equations is enlarged to include the conformal transformations.

The regularized total four-momentum can be defined as
\begin{equation}
P_\mu =\int_{\tilde\Sigma} d\sigma^\nu T_{\mu\nu},
\label{19}\end{equation}                                          
where the integration of $\Theta_{\mu\nu}$ is taken over an 
adjusted hypersurface $\tilde\Sigma$ with small 
invariant holes cut out by the future light cones drawn from 
points on the world lines slightly below their intersections 
with $\tilde\Sigma$.

To gain insight into the YMW dynamics, one should find 
simultaneous solutions of Eqs. (\ref{5}), (\ref{6}), and (\ref{8}). 
At first, one solves Eq. (\ref{5}), i.e., $A_\mu$ is expressed in terms of 
$z_I^\mu$.
Since the resulting field is singular on the world lines, its insertion 
into Eqs. (\ref{6}) and (\ref{8}) brings to ultraviolet divergences. 
There are two means of tackling this difficulty, the mass 
renormaliza\-tion and the restriction to such situations that 
Eq. (\ref{8}) can be put in its trivial form $\dot Q_I=0$. 
Upon the mass renormalization, one derives an equation of motion of the
dressed quark allowing for finite self-action.
Finally, if one succeeds in solving this equation, then the 
set of dynamical equations is entirely integrated. 

A more refined approach is to use the Noether identity (\ref{18}),
implying that the equation of motion of the dressed quark is due to  
substituting a solution of the YM equations into the equation of motion 
of the bare quark, accompanied by the mass renormalization.
On the other hand, Eq. (\ref{18}) expresses the local energy-momentum 
balance of the whole system. 

\section{Ansatz}
It is easily seen that the coefficients for highest derivatives in 
Eq. (\ref{5}) coincide with those in Maxwell equations.
Thus the characteristic cones are identical in both theories.

The retarded signal is of primary importance for every 4D 
classical field theory because it is associated with the idea of causality. 
Let us turn first to the single-quark case.
With a given point of observation $x_\mu$, such a signal carries information 
on a single point of the world line, $z_\mu^{\rm ret}$. 
Indeed, the support of the retarded function of the wave equation
\begin{equation}
D_{\rm ret}(x)=2\,\theta(x_0)\,\delta(x^2)
\label{21}\end{equation}                                          
is localized on the boundary of the future light cone, with the 
expression (\ref{21}) containing no derivatives.
The advanced function $D_{\rm adv}(x)$ reveals similar 
properties but its physical meaning is less clear. 
Other Green functions carry signals from several points 
or even from some region of the world line. 

Thus the retarded YM potential $A_\mu$ generated by a single quark may depend 
on two kinematic quantities, the four-velocity $v_\mu$ at the retarded instant 
$\tau_{\rm ret}$ and the lightlike vector $R_\mu=x_\mu-z_\mu^{\rm ret}$ drawn from the 
point of emission, $z_\mu^{\rm ret}$, to the point of observation, $x_\mu$.  

We recall elements of technique of covariant retarded quantities \cite{r}--
\cite{k}. 
Consider a plane built out of $R^\mu$ and $v^\mu$. 
A normalized vector $u^\mu$ orthogonal to $v^\mu$ and the lightlike vector 
\begin{equation}
c^\mu\equiv v^\mu+u^\mu
\label{22}\end{equation}                                          
can be drawn here. 
All this is expressible analytically as
\[v^2=-u^2=1,\quad v\cdot u=0, \quad c^2=0, \quad c\cdot v=-c\cdot u=1,\]
\[R^\mu=\rho\,c^\mu,\]
where the scalar
\begin{equation}
\rho=-u\cdot R=v\cdot R
\label{23}\end{equation}                                          
represents the distance between $z_\mu^{\rm ret}$ and $x_\mu$ in the 
reference frame with the arrow of time $v^\mu$.

From the condition $R^2=0$, one readily derives the following 
rules of differentiation: 
\begin{equation}
\partial_\mu\tau=c_\mu,
\label{24}\end{equation}                                          
\begin{equation}
\partial_\mu\rho=v_\mu+[\rho\,(a\cdot u) -1]\,c_\mu.
\label{25}\end{equation}                                          
This enables us to find derivatives of any kinematic 
quantities, for example, $\partial_\mu v_\nu =a_\nu c_\mu$. 
 
Let us further turn to the $K$-quark case.
Define the retarded invariants  
\[\rho_I\equiv R^I\cdot v^I,\quad\beta_{IJ}\equiv v^I\cdot
(R^I-R^J),\quad\gamma_{IJ}\equiv v^I\cdot v^J,\]
\begin{equation}
\Delta_{IJ}\equiv (R^I-R^J)^2=-2\,R^I\cdot R^J, 
\label{27}\end{equation}                                          
where $I,J=1,\ldots,K$, and $v_\mu^I$ is taken at $\tau_I^
{\rm ret}$. 
We have 
\[\partial_\mu\beta_{IJ}=[a^I\cdot (R^I-R^J)-1]\,c^I_\mu+
\gamma_{IJ}c^J_\mu,\] 
\[\partial_\mu\gamma_{IJ}=(a^I\cdot v^J)\,c^I_\mu+(a^J
\cdot v^I)\,c^J_\mu, \]
\begin{equation}
\partial_\mu\Delta_{IJ}=-2\,(\beta_{IJ}c^I_\mu+\beta_{JI}
c^J_\mu).
\label{28}\end{equation}                                         
Thereafter the generic retarded solution to Eq. (\ref{5}) is 
\begin{equation}
A_\mu(x)=\sum^K_{I=1}\,\sum^{N^2 -1}_{a=1} T_a\,(v_\mu^I\,
f^{aI}+R_\mu^I\,h^{aI}) ,
\label{29}\end{equation}                                          
where the sought functions $f^{aI}$ and $h^{aI}$ may depend on 
$\rho_I$, $\beta_{IJ}$, $\gamma_{IJ}$, and $\Delta_{IJ}$ 
\cite{kos}.

The expression (\ref{29}) is inserted in Eq. (\ref{5}) and the differentiations 
are made by means of Eqs. (\ref{24}), (\ref{25}), and (\ref{28}). 
One gets the expressions in which it is necessary to equate to zero the 
coefficients for the linearly independent vectors $c^I_\mu$, $v^I_\mu$, 
and $a^I_\mu$, as well as for each color basis element $T_a$. 
Recall that we search for solutions of the YM equations off the quark
world lines where the differentiation formulas (\ref{24}), (\ref{25}), and 
(\ref{28}) are just valid. 
If the procedure is to be self-consistent, we must separately equate to zero  
coefficients for every scalar kinematic quantity of which $f^{aI}$ and 
$h^{aI}$ are independent, e.g., scalars containing $a^I_\mu$. 

A distinctive feature of this procedure is that any supplementary 
condition on $A_\mu$ is unnecessary.
We thus arrive at a class of equivalence of solutions $A_\mu$ related by 
gauge transformations rather than a particular potential.

One should emphasize that the ansatz (\ref{29}) rests crucially on the 
following points: 
the field is massless;
the dynamics is gauge invariant; 
the signals are retarded;
the dimension of spacetime is four;
the world lines are timelike. 

Given a massive field, the support of the retarded Green function is the 
interior of the past light cone. 
Therefore the expression (\ref{29}) is no longer solution of the field 
equation.
It is clear that this scheme is unsuited for the dynamics without gauge 
invariance; the case is typified by replacing $\Box \eta_{\mu\nu}-
\partial_\mu\partial_\nu$ by $\Box \eta_{\mu\nu}$.

We recall also that, in $2n$-dimensional spacetimes with 
$n>2$, the retarded function is built out of derivatives 
of the $\delta$-function, therefore the retarded signal 
carries information on $v_\mu^{\rm ret}$ as well as $a_\mu^{\rm ret}$, 
and the like. 
It would be necessary to supplement the expression (\ref{29}) by 
appropriate kinematical terms. 
As to $(2n+1)$-dimensional spacetimes, the retarded signal 
carries information on the entire history of the source 
preceeding the point $z_\mu^{\rm ret}$, and the Huygens 
principle underlying our approach turns out to be invalid. 

\section{Yang-Mills field generated by a single quark}
If the source is a single quark, then it is sufficient to consider 
the gauge group SU$(2)$.
The extension to SU$(N)$ offers no significant changes in the final results.

We specify a moving basis of the color space spanned by a triplet 
$\Gamma_1^a$, $\Gamma_2^a$, $\Gamma_3^a\equiv Q^a/\sqrt{ Q^2}$ 
(with $Q^a$ precessing around $v^\mu A_\mu^a$ in the color space)
obeying the condition of orientability
\begin{equation}
\varepsilon_{abc}\,\Gamma^a_i(\tau)\,\Gamma^b_j(\tau)=
\varepsilon_{ijk}\,\Gamma_c^k(\tau),
\label{37}\end{equation}                                      
where $\varepsilon^{abc}$ are the structure constants of 
SU$(2)$, and the condition of orthonormalization
\begin{equation}
\delta_{ab}\,\Gamma^a_i(\tau)\,\Gamma^b_j(\tau)={1\over 2}\,\delta_{ij}.
\label{38}\end{equation}                                      
These relations are equivalent to Eqs. (\ref{3}) and (\ref{4}). 

A retarded solution of Eq. (\ref{5}) is written \cite{kosy} as
\[\Gamma^a_j(\tau)={\rm const},\]
\begin{equation}
A_\mu^a=\mp\,{2i\over g}\,\Gamma^a_3\,{v_\mu\over\rho}+
\kappa\,(\Gamma^a_1\pm i\Gamma^a_2)\,R_\mu,
\label{40}\end{equation}                                       
where $\kappa$ is an arbitrary nonzero integration constant with the 
dimensionality of $({\rm length})^{-2}$.

The first term of $A_\mu^a$ is a generalized Li${\rm \acute e}$nard-Wiechert 
part of the potential.
The coefficient for $\Gamma^a_3$ is an imaginary integration constant $e_3$ 
exactly fixed by the condition 
\begin{equation}
g^2\,e_3^2 =-4
\label{42}\end{equation}                                        
assuring the compatibility of an overdetermined system of nonlinear equations 
for $h_j(\rho)$. 

From Eq. (\ref{40}), one obtains the field strength 
\begin{equation}
F=c\wedge W,
\label{44}\end{equation}                                         
\begin{equation}
W^a_\mu =\mp{2i\over g}\,\Gamma^a_3\,{V_{\mu}\over \rho^2} +
\kappa\,(\Gamma^a_1\pm i\Gamma^a_2)\,v_\mu.
\label{45}\end{equation}                                         
Here, the symbol $\wedge$ signifies the exterior product of two four-vectors, and
\begin{equation}
V_\mu=v_\mu+\rho\,[a_\mu+(a\cdot u)\,u_\mu].
\label{46}\end{equation}                                          
Notice that the linearly rising term of $A_\mu^a$ contributes to the 
field strength, hence it cannot be purely gauge.

Now, the Gauss law can be represented in its familiar form: The flux of the 
generali\-zed Li${\rm \acute e}$nard-Wiechert part of the field strength through 
any two-dimensional surface, surrounding the source with the color charge 
$Q^a$, equals $4\pi Q^a$, other terms cancel out.\footnote{Although this 
result was established in the single-quark case \cite{k}, it can be extended 
to the general $K$-quark case proceeding from Eq. (\ref{13}) and taking 
advantage of a locally adjusted hypersurface ${\tilde\Sigma}$.} 
In combination with Eqs. (\ref{40}) and (\ref{44})--(\ref{46}), this yields 
the color charge of the quark 
\begin{equation}
Q^a =\pm\,{2i\over g}\,\Gamma^a_3.
\label{41}\end{equation}                                        
We draw attention to the non-analytical dependence of $A_\mu^a$ on the
coupling $g$. 
It follows that $A_\mu^a$ involves a nonperturbative information, and this is 
a good reason for sampling it as a nontrivial background in the semiclassical 
description.

For $\kappa=0$, the condition (\ref{42}) does 
not appear, and the retarded solution is
\begin{equation}
A_\mu^a=q\,\Gamma^a_3\,{v_\mu\over\rho},
\label{43}\end{equation}                                         
with $q$ being an arbitrary constant. 
Considering the field invariants 
\[F_{\mu\nu}^a\,{}^*F^{\mu\nu}_a=0,\quad F_{\mu\nu}^a\,F^{\mu
\nu}_a={4\over{g^2\rho^4}},\]
one recognizes the configuration (\ref{40}) to be 
neither self-dual nor anti-self-dual. 
Equation (\ref{40}) describes the field of the magnetic type while 
Eq. (\ref{43}) describes the field of the electric type.

If the retarded condition is replaced by the advanced one, 
then we arrive at similar expressions for the potential and 
the field strength, the only modification is the change of sign 
for $i\Gamma_2^a$ in Eqs. (\ref{40}) and (\ref{45}) as well as 
for the term in the square brackets of Eq. (\ref{46}). 
This implies that the spacetime and color arguments of $A_\mu^
a$ are correlated in a specific fashion; under time reversal, 
associated with replacing the retarded condition by the 
advanced one, the isotropic directions in the color space
$\Gamma_1^a +i\Gamma_2^a$ and $\Gamma_1^a -i\Gamma_2^a$
interchange.

Setting a new basis of the color space
\[{\cal T}_1\equiv i\Gamma_1,\quad{\cal T}_2\equiv\Gamma_2,
\quad{\cal T}_3\equiv i\Gamma_3,\]
and considering the parameter $\kappa$ to be imaginary, one 
rearranges Eq. (\ref{40}) to the form
\[A_\mu ={\cal A}_\mu^a\,{\cal T}_a\]
with real-valued ${\cal A}_\mu^a$.
Elements of the new basis can be represented by traceless 
imaginary-valued $2\times 2$ matrices satisfying the 
commutation relations of the Lie algebra sl$(2,R)$, as
becomes clear upon specifying the abstract basis 
by Pauli matrices $\Gamma_j =\sigma_j/2$.

It may appear that the doubling of colored degrees of 
freedom is attributable to opposite color charges, as Eq. (\ref{41}) suggests. 
However, the complex-conjugate potentials (\ref{40}), being represented 
in the matrix form, are interconvertible by the gauge transformation
\[{\bar A}_\mu ={\Omega}^{-1}\,A_\mu\,{\Omega}\] 
with
\[{\Omega} ={\Omega}^{-1} =\sigma_1.\]
Thus the availability of opposite color charges is deceptive in the single 
quark case. 

To sum up, we have the retarded solutions (\ref{40}) and 
(\ref{43}) describing the YM field of two 
different phases.
The first phase is specified by the noncompact gauge group 
SL$(2,R)$ while the second by the compact group SU$(2)$. 

Let us verify that Eqs. (\ref{40}) and (\ref{43}) give 
an exhaustive collection of retarded solutions to Eq. (\ref{5}) in 
that there are no other functions $f_j(\rho)$ and $h_j(\rho)$ 
representing solutions.
The potential generated by a single quark is
\begin{equation}
A_\mu^a =\sum_{j=1}^3\,\Gamma^a_j\,(\tau)\,[f_j(\rho)\,v_\mu+
h_j(\rho)\,R_\mu].
\label{47}\end{equation}                                         
Insert it into Eq. (\ref{5}).
From the requirement of vanishing the coefficient for $a_\mu$, one obtains
\[\rho\,f'_j +f_j =0\]
which is readily integrated to yield 
\begin{equation}
f_j(\rho) = {e_j\over\rho},\quad\,e_j ={\rm const}.
\label{48}\end{equation}                                         

Substituting Eqs. (\ref{47}) and (\ref{48}) in Eq. (\ref{8}) results 
in divergent terms which cannot be removed by the standard renormalization 
procedure since there is no physical parameter of an appropriate 
dimensionality at our disposal.
This difficulty can be circumvented, if one takes all constants of 
integration in Eq. (\ref{48}) to be zero, with one exception, say, $e_3$.
Thus, 
\[{\dot\Gamma}_3^a=0.\]

There remains the possibility of arbitrary rotations of 
$\Gamma_1^a$ and $\Gamma_2^a$ around ${\Gamma}_3^a$.
Having regard to the Gauss law applied to small distances, we impose an 
additional restriction on the class of sought functions: The behavior of 
$h_j(\rho)$ should be less singular than that of $f_j(\rho)$, 
namely, $\rho\, h_j(\rho)\to 0$ as $\rho\to 0$. 
Then, due to lack of finite parameters whereby these rotations can be 
specified, we conclude that 
\[\Gamma_1^a(\tau)=\Gamma_2^a(\tau)={\rm const}.\]

Let us define two isotropic color vectors
\[\Gamma_\pm^a \equiv\Gamma_1^a\pm i\Gamma_2^a\]
which together with $\Gamma_3^a$ span a new time-independent color basis.
Equation (\ref{47}) takes the form
\begin{equation}
A_\mu^a =e_3\,\Gamma^a_3\,{v_\mu\over\rho} +(\Gamma_3^a\,h_3+
\Gamma_+^a\,h_+ +\Gamma_-^a\,h_-)\,R_\mu.
\label{49}\end{equation}                                         
Substitute Eq. (\ref{49}) in Eq. (\ref{5}) and equate to zero 
the coefficients for $v_\nu$ and $R_\nu$. 
In the latter case, we separately equate to zero the 
coefficient for $a\cdot u$ and the sum of remaining terms.
Introducing $\xi\equiv\ln\rho$ and denoting the derivative with respect to 
$\xi$ by a prime, we get
\begin{equation}
h''_3 +3h'_3 +2h_3 =0,
\label{50}\end{equation}                                         
\begin{equation}
h''_+ +(3 -2ige_3)\,h'_+ +(2- 3ige_3 -g^2e_3^2)\,h_+ =0,
\label{51}\end{equation}                                         
\begin{equation}
h''_3 +h'_3 =0,
\label{52}\end{equation}                                         
\begin{equation}
h''_+ +(1 -ige_3)\,h'_+ =0,
\label{53}\end{equation}                                         
\begin{equation}
h_+h'_- -h_-h'_+ +2ige_3h_+h_- =0,
\label{54}\end{equation}                                         
\begin{equation}
e_3\,(h'_+ +2h_+) +\exp(2\xi)\,(h'_+h_3 -h'_3h_+)
-ige_3\,[e_3+\exp(2\xi)\,h_3]\,h_+ =0,
\label{55}\end{equation}                                         
and in addition three equations resulting from Eqs. (\ref{51}), (\ref{53}), and 
(\ref{55}) by the complex conjugation and the replacement of $h_+$ by $h_-$.

We have arrived at an apparently overdetermined set of 
equations: Nine equations are used to determine three sought functions.
It can be resolved if some compatibility conditions are satisfied, and this is 
accomplished if the constants of integration take certain fixed values.

Since Eqs. (\ref{50})--(\ref{53}) and their complex-conjugate are linear, we 
are seeking the simul\-taneous solution of these equations in the form 
\[h_3\propto\exp(\lambda_3\xi),\quad h_+\propto
\exp(\lambda_+\xi),\quad h_- ={\bar h}_+.\]
We get $\lambda_3 =-2$ or $\lambda_3 =-1$ from Eq. (\ref{50}), 
and $\lambda_3 =0$ or $\lambda_3 =-1$ from Eq. (\ref{52}). 
'hus, Eqs. (\ref{50}) and (\ref{52}) are á®mpatible if
\begin{equation}
h_3 =\kappa_3\,\exp(-\xi).
\label{56}\end{equation}                                         

One can find next $\lambda_+=-2+ige_3$ or $\lambda_+=-1+ige_3$ 
from (\ref{51}), and $\lambda_+ =0$ or $\lambda_+=-1+ige_3$ 
from (\ref{53}). 
Thus Eqs. (\ref{51}) and (\ref{53}) are á®mpatible if 
$\lambda_+=-1+ige_3$. The compatibility can also be
established in the case $\lambda_+=0$ which is achievable 
for $e_3=-2i/g$ or $e_3=-i/g$.

Let us examine further the compatibility of Eqs. (\ref{51}), 
(\ref{53}) with Eq. (\ref{54}). 
Assuming $h_+h_-\ne 0$,  we conclude from Eq. (\ref{54}) that
\[\lambda_- -\lambda_+ +2ige_3 =0.\]
This equation is satisfied identically for $\lambda_+=
{\bar\lambda_-}=-1+ige_3$, but it has no solution when $
\lambda_+ =\lambda_-=0$.
The compatibility of Eqs. (\ref{51}), (\ref{53}) and 
(\ref{54}) is also established for $\lambda_+=\lambda_-=0$ 
if
\begin{equation}
h_+\,h_- =0.
\label{57}\end{equation}                                         

We consider lastly the compatibility of Eq. (\ref{55}) with 
Eqs. (\ref{50})--(\ref{54}). 
Taking $\lambda_+=-1 +ige_3$, in combination with 
Eq. (\ref{56}), we obtain $e_3h_+=0$, while the 
complex-conjugate equation yields $e_3h_- =0$. 
This implies either $e_3=0$, with reducing the potential 
(\ref{49}) to the form
\begin{equation}
A_\mu^a =\gamma^a\,{R_\mu\over\rho},
\label{58}\end{equation}                                         
where $\gamma^a$ is an arbitrarily color vector, or $h_+=h_-=0$ resulting in 
\begin{equation}
A_\mu^a =\Gamma^a_3\,{e_3\,v_\mu +\kappa_3 R_\mu\over\rho}.
\label{59}\end{equation}                                         
Recall that $R_\mu/\rho=\partial_\mu\tau$, so that the 
potential (\ref{58}) is purely gauge while the expression
(\ref{59}) differs from Eq. (\ref{43}) by a gauge term.

For $\lambda_+=\lambda_-=0$, Eq. (\ref{55}) becomes 
\begin{equation}
[e_3\,(2 -ige_3) +\exp(2\xi)\,h_3]\,h_+ =0.
\label{60}\end{equation}                                         
If $e_3 =-2i/g$, then Eq. (\ref{60}) reduces to $h_+h_3 =0$, 
 nd the complex-conjugate equation is $h_-\,h_3 =0$. 
This provides two possibilities. 
First, $h_+ =h_-=0$, resulting in a purely gauge potential, 
Eq. (\ref{58}). 
Second, $h_3 =0$, which is allowable for $\kappa_3 =0$, and 
taking into account Eq. (\ref{57}), one arrives at the expression (\ref{40}).

In the case $e_3=-i/g$, Eq. (\ref{60}) is satisfied only for $h_+=0$. 
With the corresponding result for the complex-conjugate 
equation, namely, $h_-=0$, we return to the potential (\ref{58}).

So, the compatibility of all the equations is established if 
the relation (\ref{57}) holds.
In the case $h_+ =h_- =0$, there is no constraint on the 
parameter $e_3$ which yields Eq. (\ref{43}). 
On the assumption of vanishing only $h_-$ (or only $h_+$), one should equate 
$e_3$ to $-2i/g$ (or $2i/g$), and this results in Eq. (\ref{40}). 
This completes our justification of the uniqueness of the retarded solutions 
(\ref{40}) and (\ref{43}).

\section{Yang-Mills field generated by two quarks}
A detailed procedure of obtaining exact retarded solutions of 
Eq. (\ref{5}) with the source composed of two quarks, starting 
from the ansatz (\ref{29}), was given in \cite{kos}.
We thus dwell on analytical and geometrical features of these solutions.

We adopt SU$(3)$, the minimal group whereby the retarded field generated by 
two bound quarks is constructed.
The point is that the field of a bound quark occupies individually some 
SL$(2,R)$ cell of the color space while SL$(3,C)$ contains two such cells. 

One usually realizes su$(3)$ with the aid of the Gell-Mann matrices
$T_a=\lambda_a/2$. 
It is more convenient, however, for our purposes to use an overcomplete color 
basis spanned by the nonet of $3\times 3$ 
matrices including three diagonal matrices
\begin{eqnarray}                                 
H_1\equiv\frac{1}{2}\,(\lambda_3+{\lambda_8\over\sqrt 3})={1
\over 3}\pmatrix{2&0&0\cr 0&-1&0\cr0&0&-1\cr},
\label{61}\end{eqnarray}                                          
\begin{eqnarray}                                 
H_2\equiv-\frac{1}{2}\,(\lambda_3-{\lambda_8\over\sqrt 3})=
{1\over 3}\pmatrix{-1&0&0\cr 0&2&0\cr 0&0&-1\cr},
\label{62}\end{eqnarray}                                          
\begin{eqnarray}                                 
H_3\equiv-{\lambda_8\over\sqrt 3} ={1\over 3}\pmatrix{-1&0&0
\cr 0&-1&0\cr 0&0&2\cr},
\label{63}\end{eqnarray}                                          
which are related by 
\[\sum_{n =1}^3\,H_n =0,\]
and six raising and lowering matrices
\begin{eqnarray}
E_{12}^+\equiv\frac{1}{2}\,(\lambda_1+i\lambda_2)=\pmatrix
{0&1&0\cr 0&0&0\cr 0&0&0\cr},\nonumber\\E_{12}^-\equiv E_{21}^
+\equiv\frac{1}{2}\,(\lambda_1-i\lambda_2)=\pmatrix{0&0&0\cr
1&0&0\cr 0&0&0\cr},
\label{64}\end{eqnarray}                                          
\begin{eqnarray}
E^+_{13}\equiv\frac{1}{2}\,(\lambda_4+i\lambda_5)=\pmatrix
{0&0&1\cr 0&0&0\cr 0&0&0\cr},\nonumber\\E^-_{13}\equiv E^+_{
31}\equiv \frac{1}{2}\,(\lambda_4-i\lambda_5)=\pmatrix{0&0&0
\cr 0&0&0\cr 1&0&0\cr},
\label{65}\end{eqnarray}                                          
\begin{eqnarray}
E^+_{23}\equiv\frac{1}{2}\,(\lambda_6+i\lambda_7)=\pmatrix
{0&0&0\cr 0&0&1\cr 0&0&0\cr},\nonumber\\E^-_{23}\equiv E^+_{
32}\equiv \frac{1}{2}\,(\lambda_6-i\lambda_7)=\pmatrix{0&0&0
\cr 0&0&0\cr 0&1&0\cr}.
\label{66}\end{eqnarray}                                          

Given this color basis, three retarded solutions are 
\begin{equation}
A_\mu^{(1)} =\mp{2i\over g}\,(H_1\,{v_\mu^1\over\rho_1}+
H_2\,{v_\mu^2\over\rho_2})
+\kappa\,(E_{13}^\pm\,R^1_\mu + E_{23}^\pm\,R^2_\mu)\,\delta(
R^1\cdot R^2).
\label{67}\end{equation}                                          
\begin{equation}
A_\mu^{(2)} =\mp{2i\over g}\,(H_3\,{v_\mu^1\over\rho_1}+
H_1\,{v_\mu^2\over\rho_2})
+\kappa\,(E_{32}^\pm\,R^1_\mu+E_{12}^\pm\,R^2_\mu)\,\delta(R^1
\cdot R^2).
\label{68}\end{equation}                                          
\begin{equation}
A_\mu^{(3)} =\mp{2i\over g}\,(H_2\,{v_\mu^1\over\rho_1}+
H_3\,{v_\mu^2\over\rho_2})
+\kappa\,(E_{21}^\pm\,R^1_\mu+E_{31}^\pm
\,R^2_\mu)\,\delta(R^1\cdot R^2).
\label{69}\end{equation}                                          
They represent actually the same YM field related by the gauge transformations 
\[A^{(j)}_\mu={\Omega}_{1j}^{-1}\,A^{(1)}_\mu\,{\Omega}_{1j}\]
with
\[{\Omega}_{12} =\pmatrix{0&0&1\cr 1&0&0\cr 0&1&0\cr},\quad 
{\Omega}_{13} ={\Omega}^{-1}_{12}=\pmatrix{0&1&0\cr 0&0&1\cr
1&0&0\cr}.\]
Therefore, only one of Eqs. (\ref{67})--(\ref{69}), say, 
Eq. (\ref{67}), will thereafter be referred to.

Taking into account the Gauss law, one finds the color 
charge of $I$th quark 
\begin{equation}
Q_I =\mp\,\frac{2i}{g}\,H_I.
\label{70}\end{equation}                                          
Thus the solutions suggest the existence of two-quark 
systems with the total color charges
\begin{equation}
Q_{(+)}={2i\over g}\,(H_1+H_2)\quad{\rm and}\quad Q_{(-)}=-{2i\over g}
\,(H_1+H_2).
\label{71}\end{equation}                                           
Therein lies the most outstanding distinction from the single-quark case
where the complex-conjugate potentials, being interconvertible by gauge 
transformations, originate from the same source.
As for the present case, it is impossible to convert the complex-conjugate 
solutions to one another since there exists a nonzero field invariant
\[C_3 ={\rm tr}\,(F_{\lambda\mu}\,F^\mu_{\hskip1.5mm\nu}\,F^{\nu\lambda})\]
which is of different signs for the complex-conjugate solutions.
So, we have two different field configurations generated by 
two sources with the total color charges $Q_{(+)}$ and $Q_{(-)}$.
It follows from Eq. (\ref{71}) that no colorless two-quark cluster 
is feasible on the classical level. 

Let us turn to the spacetime dependence of the solutions.
It was found in \cite{kos} that $f^{iI}$ and $h^{iI}$ are independent of 
$\beta_{12}$ and $\gamma_{12}$, whereas $\Delta_{12}$ is shown by the factor 
$2\,\delta(\Delta_{12})=\delta(R_1\cdot R_2)$.

Since $R_1$ and $R_2$ are lightlike, the equality $R_1\cdot 
R_2=0$ is ensured only for collinear $R_1$ and $R_2$. 
In view of the factor $\delta(R^1\cdot R^2)$, this means that the linearly 
rising term of $A_\mu$ is localized on the enveloping surface of 
two families of rays
\[x_\mu=z_\mu^1(\tau)+\theta(\sigma)\,n_\mu\sigma,\]
\[n_\mu =z^1_\mu-z^2_\mu,\quad n^2=0,\quad n_0>0,\]
and
\[x_\mu=z_\mu^2(\tau)+\theta(\sigma)\,m_\mu\sigma,\]
\[m_\mu=z^2_\mu-z^1_\mu,\quad m^2 =0,\quad m_0>0,\]
parametrized by $\tau$ and $\sigma$.       
At the intersection of a spacelike hyperplane with this surface, two 
fragments of a curve arise. 
Thus the force lines of the YM field corresponding to the linearly 
rising term of $A_\mu$ are squeezed to a string. 
This is not a finite string joining two quarks. 
What we have now are two half-infinite strings which begin at the positions 
of the quarks and go outward at different sides to spatial infinity. 

By the construction of the enveloping surfaces, they are ruled surfaces.
Having lightlike rulings and timelike directrices (the world lines), 
we find warped worldsheets of the strings.

There exists also an alternative solution identical to Eq. (\ref{67}) in every 
respect except the linearly rising term that is stripped of the factor 
$\delta(R^1\cdot R^2)$,
\begin{equation}
A_\mu=\mp{2i\over g}\,(H_1\,{v_\mu^1\over\rho_1}+g\,
\kappa\,E_{13}^\pm\,R^1_\mu)
\mp\,{2i\over g}\,(H_2\,{v_\mu^2\over\rho_2}+g\,\kappa\,
E_{23}^\pm\,R^2_\mu).
\label{72}\end{equation}                                        
Now the linearly rising terms describe the force lines 
distributed over all directions of space.

The solution (\ref{72}) is truly non-Abelian because 
\[[A_\mu,A_\nu]\ne 0,\quad [A_\mu,F^{\mu\nu}]\ne 0.\]
How can the nonlinearity of the YM equations be compatible with that 
$A_\mu$ is the sum of two single-quark potentials?
Equation (\ref{72}) does combine two such terms, yet making no hint about the 
plausibility to represent the solution as an arbitrary superposition of them.
If either of terms with some coefficient different from 1 is added 
to another, no new solution arises. 

It should be realized that Eq. (\ref{72}) describes
the YM fields generated by a system of two {\it bound} quarks.
Would the field be generated by two free quarks, the sign of 
the color charge of any quark might be subjected to variation 
regardless of the sign of the color charge of other quark. 
In the present case, however, changing the signs for the first 
one-quark term and leaving intact the signs for the second one-quark term, we
arrive at the expression that is no longer solution as it is clear 
from that $E_{13}^\pm$ and $E_{23}^\mp$ do not commute.
One may change the signs only simultaneously for both terms.
This correlation of signs is precisely the evidence that we 
are dealing with bound quarks. 

For $\kappa=0$, a retarded solution is a superposition of two
single-quark potentials (\ref{43}),
\begin{equation}
A_\mu=\sum_{I=1}^2\sum_{n=1}^3\,e_I^n\,H_n\,{v_\mu^I\over
\rho_I}.
\label{73}\end{equation}                                   

The solutions (\ref{67})--(\ref{69}) and (\ref{72}) become real-valued
with respect to the color basis
\[{\cal T}_1\equiv i\frac{\lambda_1}{2},\quad{\cal T}_2\equiv
\frac{\lambda_2}{2},\quad{\cal T}_3\equiv i\frac{\lambda_3}
{2},\quad{\cal T}_4\equiv i\frac{\lambda_4}{2},\]
\[{\cal T}_5\equiv\frac{\lambda_5}{2},\quad{\cal T}_6\equiv i
\frac{\lambda_6}{2},\quad{\cal T}_7\equiv\frac{\lambda_7}{2},
\quad{\cal T}_8\equiv i\frac{\lambda_8}{2},\]
or else
\[{\cal H}_n\equiv i\,H_n,\qquad{\cal E}_{mn}^\pm\equiv i\,
E_{mn}^\pm.\]
With reference to the explicit form of $H_n$ and $E_{mn}^\pm$, 
Eqs. (\ref{61})--(\ref{66}), one finds that 
${\cal T}_n$ are traceless imaginary $3\times 3$ matrices
satisfying the commutation relations of the Lie algebra sl$(3,R)$. 
'hus the gauge symmetry of the solutions 
(\ref{67})--(\ref{69}) and (\ref{72}) is SL$(3,R)$.

When either of two quarks, say, the first, is elimitated, then 
Eq. (\ref{68}) acquires the form
\[A_\mu=A_\mu'+A_\mu'',\]
\begin{equation}
A_\mu'=\mp\,{i\over g}\,\lambda_3\,{v_\mu\over\rho}+\kappa\,
E^{12}_\pm\,R_\mu,\quad A_\mu''=\mp\,{i\over g}\,{\lambda_8
\over\sqrt 3}\,{v_\mu\over\rho}.
\label{74}\end{equation}                                          
$A_\mu'$ is the single-quark solution (\ref{40}) while $A_\mu''$ is an Abelian 
term, decoupled from $A_\mu'$ since $\lambda_8$ commutes with $\lambda_3$ and 
$E^\pm_{12}$. 
'he adequacy of the gauge group SU$(2)$ in the single-quark case is thus 
confirmed; the non-Abelian piece of the solution is built out of color 
vectors forming the Lie algebra su$(2)$. 
The field invariant $C_3$ is zero for both Eqs. (\ref{40}) and (\ref{74}).

It is no great surprise that SL$(2,C)$ stands out against SL$(N,C),\,N>2$, 
in the single-quark case.
The metrical structure of the base embodied in the Lorentz group SL$(2,C)$ is 
all that should be mapped by the future light cone into the fiber, so that the 
color space SL$(2,C)$ is the only exact image.
Based on SO$(N)$ or Sp$(N)$ as the starting point, one reaches the same 
solution since su$(2)\sim {\rm so}(3)\sim{\rm sp}(1)$, and 
sl$(2,R)\sim{\rm su}(1,1)\sim {\rm so}(2,1)\sim{\rm sp}(1,R)$ \cite{barr}. 

On the other hand, given SO$(N)$ or Sp$(N)$ in the two-quark 
case, we have other results as opposed to SU$(N)$.
Both the so$(4,C)$ and so$(5,C)$ color spaces are suitable for 
an accommodation of two ``elementary'' color cells so$(3,C)\sim {\rm sl}(2,C)$.
But so$(4,C)$ is not semisimple, and the Cartan-Killing metric is singular 
here.
As for so$(5,C)$, it is isomorphic to sp$(2,C)$, and we envisage two 
alternatives in the description of the color space in the 
two-quark case, either sl$(3,C)$ or so$(5,C)\sim{\rm sp}(2,C)$.  

\section{Yang-Mills field generated by several quarks}
The discussion of solutions of the YM equations with the source composed of 
$K$ quarks echoes in many respects that in the two-quark case. 
We adopt now the gauge group SU$(N)$ with sufficiently large $N$, at least 
$N\ge K+1$, to allow an accommodation of all $K$ quarks.
\subsection[section]{Solutions}
We use the Cartan-Weyl basis of the Lie algebra su$(N)$ spanned by the set of 
$N^2$ matrices which includes $N$ diagonal elements $H_n$, the Cartan 
subalgebra,
\[(H_n)_{AB}\equiv\delta_{An}\,\delta_{Bn}-N^{-1}\,\delta_
{AB},\]
satisfying 
\[\sum_{n=1}^N H_n =0,\] 
and $N^2-N$ raising and lowering elements $E_{mn}^+$ and $E_{mn}^-$, with 
$n>m$, 
\[(E^+_{mn})_{AB}\equiv\delta_{Am}\,\delta_{Bn},\quad
(E^-_{mn})_{AB}\equiv\delta_{Bm}\,\delta_{An}.\]
Here,  $m,\,n,\,A,\,B = 1,\dots,N$.
The nontrivial commutators are as follows 
\begin{equation}
[H_m,\,E^\pm_{mn}]=\pm\,E^\pm_{mn},
\label{751}\end{equation}                                    
\begin{equation}
[E^+_{mn},\,E^-_{mn}]=H_m -H_n,
\label{752}\end{equation}                                    
\begin{equation}
[E^\pm_{kl},\,E^\pm_{lm}] =\pm\,E^\pm_{km}.
\label{75}\end{equation}                                      

With these commutation relations, one can ascertain that Eq. (\ref{5}) 
is satisfied by 
\begin{equation}
A_\mu=\mp\,{2i\over g}\sum_{I=1}^K\,[\,H_I\,{v_\mu^I\over
\rho_I}+g\,\kappa\,E^\pm_{I\,K+1}\,R_\mu^I\,\prod_{I=1}^{K-1}
\delta(R^K\cdot R^I)\,].
\label{76}\end{equation}                                      
There exist $C^K_N$ solutions of this type.
Consider the transformation $A_\mu\to{\Omega}^{-1}A_\mu{\Omega}$ with
\[\Omega=E^-_{1\,N}+\sum^{N-1}_{I=1} E^+_{I\,I+1}=\pmatrix{
0&.&.&.&{}&{}&0&1\cr 1&0&{}&{}&{}&{}&{}&{}
\cr {}&1&0&{}&{}&{}&{}&{}\cr {}&{}&.&.&{}&{}&{}&{}\cr
{}&{}&{}&.&.&{}&{}&{}\cr {}&{}&{}&{}&.&.&{}&{}\cr
{}&{}&{}&{}&{}&{}&{}&{}\cr}\]
and
\[\Omega^{-1} = E^+_{1\,N}+\sum^{N-1}_{I=1} E^-_{I\,I+1}=
\pmatrix{0&1&{}&{}&{}&{}&{}&{}\cr {}&0&1&{}&{}&{}&{}&{}\cr
{}&{}&.&.&{}&{}&{}&{}\cr {}&{}&{}&.&.&{}&{}&{}\cr
{}&{}&{}&{}&.&.&{}&{}\cr {}&{}&{}&{}&{}&{}&0&1\cr
1&0&.&.&.&{}&{}&0\cr}\]
It increases each index of $H_I$ and $E^\pm_{I\,K+1}$ by one, 
and the transformed $A_\mu$ turns out to be a new solution. 
Other solutions can be obtained by  repetitions of this gauge transformation.
[A similar situation already encountered in the two-quark case where the 
solutions (\ref{67})--(\ref{69}) were shown to convert to each other by 
transformations of this sort].

The solution (\ref{76}) describes a background field generated by 
$K$ quarks which form some $K$-quark cluster.
The color charge of $I$th quark [cf. Eq. (\ref{70})] is 
\[Q_I =\mp\,{2i\over g}\,H_I.\]
The total color charge of such a cluster [cf. Eq. (\ref{71})] is either 
\[ Q_{(+)}=\frac{2i}{g}\,\sum_{I=1}^K\,H_I\quad{\rm or}\quad Q_{(-)}=-
\frac{2i}{g}\,\sum_{I=1}^K\,H_I.\]

There are also solutions describing background fields generated by several 
clusters. 
Each cluster is defined by the condition that the signs of the 
color charges are simultaneously either $+$ or $-$ for 
every quark of the cluster whereas relative signs of the total 
color charges of the clusters are arbitrary. 
For example, the potential generated by two two-quark 
clusters is $A_\mu=A_\mu^1\pm A_\mu^2$  where $A_\mu^j$ is the potential 
generated by the $j$th cluster,
\begin{eqnarray}
A_\mu^1=\pm\,{2i\over g}\sum_{I=2}^3\,[\,H_I\,{v_\mu^I\over
\rho_I}+g\,\kappa\,E^\pm_{1\,I}\,R_\mu^I\,\delta(R^2\cdot R^3)
\,],\nonumber\\ A_\mu^2=\pm\,{2i\over g}\sum_{I=5}^6\,[\,H_I\,
{v_\mu^I\over\rho_I}+g\,\kappa\,E^\pm_{4\,I}\,R_\mu^I\,\delta
(R^5\cdot R^6)\,].
\label{77}\end{eqnarray}                                          
Omitting $\delta(R^I\cdot R^J)$ in Eqs. (\ref{76}) and (\ref{77}) gives 
alternative solutions [cf. Eq. (\ref{72})]. 

At last, there exist solutions describing YM fields of free quarks with the 
color charges
\[Q_I=\pm\,\frac{i}{g}\,(\,H_{I+1}-H_I).\] 
For example, the YM field of two free quarks (labeled by numbers 1 and 3) is 
\begin{equation} 
A_\mu=\pm\,[\,{i\over g}\,(\,H_2-H_1)\,{v_\mu^1\over
\rho_1}+\kappa\,E^\pm_{1\,2}\,R_\mu^1]
\pm\,[\,{i\over g}\,(H_4-H_3)
\,{v_\mu^3\over\rho_3}+\kappa\,E^\pm_{3\,4}\,R_\mu^3\,].
\label{78}\end{equation}                                          

The gauge transformation $A_\mu\to{\Omega}^{-1}A_\mu{\Omega}$ 
with 
\[{\Omega}={\Omega}^{-1}={N-2\over N}\,{\bf 1}+\sum_{I=3}^N
\,H_I+E^+_{1\,2}+E^-_{1\,2}
=\pmatrix{0&1&{}&{}&{}&{}&{}\cr
1&0&{}&{}&{}&{}&{}\cr {}&{}&1&{}&{}&{}&{}\cr 
{}&{}&{}&1&{}&{}&{}\cr {}&{}&{}&{}&.&{}&{}\cr
{}&{}&{}&{}&{}&.&{}\cr {}&{}&{}&{}&{}&{}&.\cr}\]
changes the $\pm$ signs of the first square bracket of 
Eq. (\ref{78}) while the signs of the second remains invariable. 
It is easy to recognize this gauge transformation as that 
rendering the potential complex-conjugate in the single-quark case.  
Thus the color charge of the free quark is determined modulo 
$\exp(i\pi n)$.

There is a number of ways to separate a given $K$-quark system into 
groups of clusters of a certain quark content and free quarks.
We interpret them as {\it scenarios of hadronization}.

We now look at the symmetry of these solutions.
One can define $(K+1)^2$ traceless imaginary matrices 
${\cal H}_n$ and ${\cal E}_{mn}^\pm$ as follows:
\[{\cal H}_n\equiv i\,H_n,\quad
{\cal E}_{mn}^\pm\equiv i\,E_{mn}^\pm\] 
which are elements of the Lie algebra sl$(K+1,R)$.
Thereafter, every solution above becomes real valued with respect to this 
basis.
The solutions constructed from $M^2$ such elements obeying 
the closed set of commutation relations are invariant under 
SL$(M,R)$, $M\le K+1$.

In particular, the YM field generated by a two-quark cluster (meson) 
is invariant under SL$(3,R)$ and that of a three-quark cluster (barion) is 
invariant under SL$(4,R)$.
Since SL$(3,R)$ is a subgroup of SL$(4,R)$, the YM field of every hadron is 
specified by the gauge group SL$(4,R)$. 
This symmetry is independent of $N$ and is retained in the limit $N\to\infty$.

For $\kappa =0$, the YM equations linearize, and one gets an Abelian solution 
\begin{equation}
A_\mu=\sum_{I=1}^K \sum_{n=1}^N\,e_I^n\,H_n\,{v_\mu^I\over\rho_I},
\label{79}\end{equation}                                    
where $e_I^n$ are arbitrary parameters. 
The gauge symmetry of this solution is SU$(N)$.

We have obtained two types of solutions corresponding to two phases of matter.
The YM background of the first phase is invariant under 
the noncompact gauge group SL$(K+1,R)$ while that of the 
second phase is invariant under the compact gauge group SU$(N)$. 

\subsection{Energetical considerations}
It follows from the trace relations
\begin{equation}
{\rm tr}\,(H_l\,E^\pm_{mn})=0
\label{tr}\end{equation}                                    
that the linearly rising term of $A_\mu$ does not contribute to the color 
force 
\begin{equation}
f^\mu_I=v_\nu^I\,{\rm tr}\,\bigl[Q_I\, F^{\mu\nu}(z_I)\bigr].
\label{f}\end{equation}                                    
Thus, the well-known mechanism of quark binding by a constant force lacks 
support from the exact solutions of the classical YM equations.

A surprising thing is that the linearly rising term of $A_\mu$, 
while ensuring nonzero contribution to the field strength 
$F_{\mu\nu}$, results in no force.
An explanation is simple.
Equation (\ref{f}) includes the scalar product of two color vectors 
$F_{\mu\nu}$ and $Q_I$ which are not arbitrary; they come from the exact solutions and 
turn out to be orthogonal to each other.

One can find from Eq. (\ref{tr}) and
\begin{equation}
{\rm tr}\,(E^\pm_{mn}\,E^\pm_{mn})=0
\label{tre}\end{equation}                                    
that the linearly rising term of $A_\mu$ does not contribute to color singlets 
altogether.
This is because it depends on either $E_{mn}^+$ or $E_{mn}^-$, not both.
A more fundamental reason is the conformal invariance of the 4D 
classical YM theory which implies that the parameter $\kappa$ violating 
the scale symmetry cannot manifest itself in observables.

The energy-momentum tensor can be splitted into 
\[\Theta_{\mu\nu}=\sum_{I=1}^K\,\Theta_{\mu\nu}^I+
\Theta_{\mu\nu}^{\rm int},\] 
where $\Theta_{\mu\nu}^I$ is the self-action term containing the contribution 
of the YM field generated by $I$th quark and $\Theta_{\mu\nu}^{int}$ is the 
interaction part comprised of mixed contributions. 

The four-momentum $P_\mu$ defined by Eq. (\ref{19}) contains divergent terms 
due to $\Theta_{\mu\nu}^I$'s.
If the solution is invariant under SL$(K+1,R)$, the self-energy of each 
quark is negative definite.
This suggests that such backgrounds are most favorable at zero temperature. 
It is the energetical advantage that enables attributing them 
to the gluon vacuum in the cold world.

The self-energy is positive-definite for the solutions invariant under SU$(N)$.
These solutions seem to be related to the hot phase.
[One should mention also the configurations \cite{jack} invariant 
under SU$(N)$ with an energy lower than that of the Coulomb solution].

However, it is beyond the scope of this work to review a temperature-dependent 
version of the YMW theory.
We merely drew a parallel to the Yang-Mills-Higgs (YMH) theory where two 
classical solutions with different symmetries also exist. 
The solution with bro\-ken symmetry being stable and energetically favorable 
corresponds to the cold phase. 
Although the solution with unbroken symmetry is unstable (thus bearing no 
relation to physical world), its availability on the fundamental level 
motives the quest of a phase with such a symmetry.   
 
The present situation contrasts with that of the YMH theory in three aspects.
First, there is no spontaneous symmetry breakdown.
We deal with solutions invariant under two different real forms of the 
complex group SL$(N,C)$.
The occurrence of the solution invariant under a noncompact group different 
from the initial one is a new field-theoretic phenomenon referred to as 
the {\it spontaneous symmetry deformation} \cite{kos}. 
The epithet spontaneous emphasizes that the scenario of hadronization is
accomplished quite accidentally.

Second, both solutions are now stable against small disturbances (see Sec. VII).

Third, the critical point $\kappa=0$ is independent of parameters 
appearing in the action, as opposed to the YMH theory where the spontaneous 
symmetry breakdown is directly related to parameters controlling the 
convexity of the Higgs potential.

Let us turn to the self-action problem.
We follow the basic Teitelboim pattern \cite{teit} developed in the 
Maxwell-Lorentz theory.
The SU$(N)$ phase is treated with no noticeable distinctions from 
electrodynamics; one needs simply to substitute everywhere $e^2$ by $Q_I^2$. 

A completely different situation arises in the cold phase.
After the mass renormalization, the negative definiteness of 
the field energy does not disappear without leaving a trace.
It reveals itself in the ``wrong" sign  of the radiation energy of  
accelerated quarks \cite{kosy,k}, 
\begin{equation}
\frac{dE_I}{d\tau_I}=\frac{2}{3}\,\vert\,{\rm tr}\,Q^2_I\,\vert\,a^2_I<0.
\label{80}\end{equation}                                          
Thus the self-action of a quark in the cold phase is such that 
the flux of energy is directed inward the source.
This feature of the self-action is unrelated to the boundary 
condition; replacing the retarded condition by the advanced one
leaves the direction of the flux intact.

To gain better understanding of this point, one should derive 
the equation of motion of a dressed quark.
Taking into account Eqs. (\ref{16})--(\ref{18}), one can find \cite{kosy,k} 
\begin{equation}
m_I\,[a^I_\mu+\tau_0\,(\dot a^I_\mu+v^I_\mu\, a_I^2)]=
v^\nu_I\,{\rm tr}[Q_I F_{\mu\nu}(z_I)].
\label{81}\end{equation}                                          
Here, $m_I$ is the renormalized mass of $I$th quark, 
$F_{\mu\nu}$ is the field of all other quarks at the 
position of $I$th quark $z_I^\mu$, and 
\[\tau_0\equiv\frac{2}{3m_I}\,\vert\,{\rm tr}\,Q^2_I\,\vert.\]

A similar parameter in electrodynamics $\tau_0=2e^2/3mc^2\approx 10^{-13}$ 
cm is related to the classical radius of electron.
Every effect of the scale $\tau_0$ is neglected there, keeping in mind that 
quantum phenomena come into play already at the range of the Compton 
wavelength of electron $\lambda=\hbar/mc\approx 10^{-11}$ cm.
On the contrary, the classical radius of quark in the cold phase far exceeds 
its wavelength, $\tau_0\gg\lambda$, since, in the semiclassical treatment, 
the coupling $g$ is held to be much less than 1, and therefore
\begin{equation}
\vert\,{\rm tr}\,Q^2_I\,\vert=\frac{4}{g^2}(1-\frac{1}{N})\gg 1.
\label{811}\end{equation}                                          

Equation (\ref{81}) can be rewritten as
\begin{equation}
\stackrel{\scriptstyle v}{\bot}\,({\dot p}-f)=0,
\label{82}\end{equation}                                          
where the operator
\begin{equation}
\stackrel{\scriptstyle v}{\bot}\,\equiv\, {\bf 1}-
\frac{v \otimes v }{v^2}
\label{32}\end{equation}                                          
projects vectors on a hyperplane normal to the four-velocity $v^\mu$, and
\begin{equation}
p^\mu=m\,(v^\mu+\tau_0\,a^\mu).
\label{83}\end{equation}                                          
Note that Eq. (\ref{82}) is Newton's second law governing the 
behavior of an object specified by its point of location $z^\mu$ and 
four-momentum $p^\mu$ defined by Eq. (\ref{83}).
We call this object the dressed quark, or the {\it color complex} 
\cite{k}, keeping in mind the field-mechanical origin of $p^\mu$.  

The invariance of the action (\ref{2}) under the local reparametrizations 
\[\delta \tau=\epsilon,\quad \delta z^\mu=v^\mu\epsilon\] 
endows the equation of motion of a bare quark with the factor $\bot$. 
The regularization of $P_\mu$ indicated in Sec. II ensures that such an 
invariance remains valid after the mass renor\-malization.
Thus the form of the dynamical equations of bare and dressed quarks is the 
same, only their four-momenta $p^\mu$ are distinct in the dependence on 
kinematical variables.

The complex is unaffected by the {\it radiation reaction}; the Newtonian 
behavior of this object implies that only external force acts on it.  
The hallmark of the complex is not the evolution law, which is not uncommon, 
but the indefiniteness of $p^0$, as Eq. (\ref{83}) suggests.
 
On the other hand, Eq. (\ref{81}) expresses the local 
four-momentum balance: The increment of the complex four-momentum $dp_\mu^I$ 
originates from the total effect of all other quarks $v^\nu_I\,{\rm tr}\,
[Q_I F_{\mu\nu}(z_I)]\,d\tau_I$ and the absorbed four-momentum, 
$\tau_0\,v^I_\mu\,a_I^2\,d\tau_I$ \cite{k}. 
The greater the acceleration (determined by the total effect of other 
quarks) the greater the absorption. 

A more familiar viewpoint is that Eq. (\ref{81}) describes the evolution of an 
object with the four-momentum $p^\mu=m\,v^\mu$, visualized as a point particle. 
The behavior of such a `particle' is beyond the control of the Newton 
second law. 
Departures from the Newtonian behavior are commonly 
attributed to the radiation reaction.
This is a source of many paradoxes \cite{k}.

For a vanishing RHS of Eq. (\ref{81}), one gets the solution
\[v^\mu=\{\cosh(C+De^{-\tau_/\tau_0}),\,{\bf n}\sinh(C+
De^{-\tau_/\tau_0})\},\]
where $C$ and $D$ are constants, and {\bf n} is a fixed unite vector.
Thus, in the absence of the external force, the absorption of energy is 
exponentially decreasing in time, and the motion of the color complex 
asymptotically approaches Galileo's inertial regime.
The solution describes a straight world line when the asymptotical condition 
$a_\mu\to 0$, $\tau\to -\infty$ is imposed. 

We further turn to the interquark forces. 
In the cold phase, like color charges attract and unlike ones repel.
However, it follows from the trace relations
\begin{equation}
{\rm tr}\,(H_{J+1}-H_J)^2=2,\quad{\rm tr}\,(H_{J+1} -H_J)\,H_I=0
\label{84}\end{equation}                                          
that a free quark, while experiencing the self-action, does not 
act on other quarks.

We have seen that, in the cold phase, each quark individually occupies some 
sl$(2,R)$ cell.
Neither of two backgrounds generated by different quarks may be 
contained in the same sl$(2,R)$.
This is similar to the Pauli blocking principle.
Just as a cell of volume $h^3$ in the phase space might be 
occupied by at most one fermion with a definite spin polarization, so 
any sl$(2,R)$ cell is intended for a background of only one quark.
Choosing SO$(N)$ or Sp$(N)$, instead of SU$(N)$, one singles out the same 
color sell so$(2,1)\sim{\rm sp}(1,R)\sim{\rm sl}(2,R)$.

By contrast, in the hot phase, assuming the total color charge of quarks 
in a given plasma lump to be zero, the parameters $e_I^n$ in Eq. (\ref{79}) 
are to be appropriately fitted.
Then the most energetically advantageous field configuration is such that 
the color charges of quarks are lined up into a fixed color direction,  
thereby reducing SU$(N)$ to SU$(2)$.
This bears some resemblance to the Bose-Einstein condensation in the color 
space.

Thus the ``color Pauli principle'' preventing a body of $K+1$ color cells 
against shrinkages is an evidence of that the large-$N$ limit is adequate to 
the cold phase description, whereas the ``color Bose-Einstein condensation'' 
suggests the sufficiency of SU$(2)$ for the hot phase.
 
Consider $N\to\infty$ limit in the cold phase, assuming the coupling 
$g$ to be fixed. 
(Note that the factorization condition is therewith assured).
The relations
\begin{equation}
{\rm tr}\,(H_I)^2=1-N^{-1} ,\quad{\rm tr}\,(H_I\,H_J)=-N^{-1}
\label{841}\end{equation}                                          
show that the color repulsion between bound quarks vanishes in this limit, 
unless the number of quarks at the given cluster is of order $N$. 
Thus $K$-quark clusters with $K=O(N)$ are unstable while any cluster of finite 
number of quarks survives as $N\to\infty$. 

This is in agreement with Witten's phenomenology \cite{wit}, where
mesons made out of quark-antiquark pairs are stable and noninteracting 
(their decay and scattering amplitudes are suppressed, respectively, as 
$1/\sqrt{N}$ and $1/N$), and barions, imagined as $N$-quark clusters, are 
unstable (the barion-barion and barion-meson vertices are respectively of 
order $N$ and 1). 

In the present context, however, barions being considered as three-quark 
clusters turn out to be stable.
The consistency with Witten's phenomenology is true for both mesons and
multiquark clusters with the number of quarks of order $N$.

Thus, the classical YMW quarks do not interact as $N\to\infty$.
The quark binding is characterized only by the correlation of signs of the 
color charges of quarks comprising a cluster.
Since the color self-action prevents the motion from the runaway regime, the 
quarks could be conceived as moving along parallel straight world lines.
This accords with an intuitive idea of the ground state of a cluster with zero 
orbital momentum.
\footnote{A similar situation occurs in the monopole dynamics.
As was shown by Manton \cite{man}, the monopoles forming a static 
multimonopole are influenced by no intermonopole forces; they are balanced 
due to exact cancellation of the repulsive magnetic YM force and the 
attractive Higgs force.} 

Switching on electromagnetic forces violates this equilibrium. 
It is possible, however, to introduce the electro\-dynamical terms of the 
action in conjunction with choosing the YM coupling $g$ such that 
\begin{equation}
\vert\,{\rm tr}\,Q^2_I\,\vert=\frac{4}{g^2}(1-\frac{1}{N})=e^2.
\label{842}\end{equation}                                          
This enables a consistent treatment of orbital motions.
It follows from Eqs. (\ref{841}) and (\ref{842}) that the centrifugal force 
is finite while the absorbed YM energy exactly compensates the radiated
electromagnetic energy, and helical world lines are no longer infrared 
troublesome.
Moreover, the resulting picture is free of ultraviolet divergences.

Unfortunately, this attractive possibility may pretend only to a toy status.
With the actual value of the elementary electric charge $e^2\approx 1/137$, 
Eq. (\ref{842}) results in $g\approx 24$ invalidating the semiclassical 
tratment.
Furthermore, the electric charge of real quarks takes two values $-e/3$ and 
$2e/3$ so that the picture of accelerated quarks emitting no energy can be 
matched with only either type of quarks (and with electroneutral two-quark 
clusters).

\section{Stability problem} 
We have found that the YMW  bound quarks are balanced being in the state 
of indifferent equilibrium.
One should then examine the spectrum of excitations about the classical 
background.
In order that a given cluster be stable, the energy of every mode
(the translation mode apart) must be positive; if the balance is upset by 
some external influence, then excitations responsible for increasing the 
energy should occur. 

For our prime interest is in the ground state of clusters where the quarks 
rest relative to each other, we consider the static background field 
generated by such quarks.
Set $B_\mu=A_\mu+b_\mu$ where $A_\mu$ is a static configuration, and $b_\mu$ 
is a small disturbance about $A_\mu$. 
As can be readily shown (see, e.g., \cite{raj}), the positivity of the
excitation energy about a given static background $A_\mu$ is tantamount 
to that the equation of excitations
\begin{equation}
{\delta^2S\over\delta A^a_\mu(x)\,\delta A^b_\nu(y)}\,b_\mu^a(x)=0
\label{85}\end{equation}                                          
has no solution exponentially increasing with time. 
Any oscillatory solution $\exp\,(i\omega_k t)$ determines a positive mode
$\varepsilon_k\propto\omega^2_k$.
(For a more extended discussion see, e.g., \cite{jr}.)

Let us show that the single-quark solution (\ref{40}) is stable against 
small field disturbances \cite{kosya}.
In the static case, when $v^\mu=(1,0,0,0)$,\,$z^\mu(\tau)=z^\mu(0)+v^\mu\tau$, 
the proper time $\tau$ is identified with the laboratory time $t$,  nd the 
retarded distance $\rho$ with the usual radius $r$. 

As is well known \cite{raj}, the classical limit $\hbar\to 0$ is equivalent to 
that of the weak coupling $g\to 0$.
Taking into account that the expression (\ref{40}) depends on $g$ as $g^{-1}$, 
we must retain only quantities of order $g^0$ in Eq. (\ref{85}).

Let us take the gauge condition $v^\mu b_\mu^a =0$ for any quark world line.
Then the color charge of the quark remains constant $\dot Q^a=0$ even with 
the presence of the excitations $b_\mu^a$. 
In the static case, this condition is reduced to 
\[b^a_0=0.\]

Among the spatial components of ${\bf b}^a$, we must separate only those which 
are orthogonal to the gauge modes.
This is guaranteed by the condition \cite{co}
\[\nabla{\bf b}^a+gf^{abc}\,{\bf B}^b\cdot{\bf b}^c=0,\]
which, in the weak coupling limit, becomes 
\begin{equation}
\nabla{\bf b}^a=0. 
\label{86}\end{equation}                                        

Putting 
\[{\bf b}={\bf b}^3\Gamma_3+{\bf b}^+\Gamma_+ +{\bf b}^-\Gamma_-,\]
and taking into account Eq. (\ref{86}), we obtain \cite{kosya} 
\begin{equation}
\Box\,{\bf b}^3=0,
\label{87}\end{equation}                                        
\begin{equation}
(\Box\mp{4\over r}\,{\partial\over\partial t}+{4\over r^2})\,
{\bf b}^\pm=0,
\label{88}\end{equation}                                         
\begin{equation}
{\bf r}\cdot{\bf b^\pm} =0.
\label{89}\end{equation}                                          

It is clear from Eq. (\ref{87}) that ${\bf b}^3$ does not violate 
the stability of the background $A_\mu^a$. 
The function ${\bf b}^-$, satisfying Eqs. (\ref{86}), (\ref{88}), and 
(\ref{89}), with oscillatory behavior in time is 
\begin{equation}
{\bf b}^-(t,{\bf r})=\int^\Lambda_0 d\omega
\sum_{l,m}\lbrace\alpha_{lm}(\omega)\,e^{-i\omega t}\,
{\bf Y}_{lm}(\theta,\phi)\,K_j(\omega r)
+\beta_{lm}(\omega)\,e^{i\omega t}\,\bigl[{\bf Y}_{lm}(\theta,
\phi)\,K_j(\omega r)\,\bigr]^*\rbrace.
\label{90}\end{equation}                                     
Here, $\Lambda$ is a frequency cutoff parameter that characterizes a boundary 
of the infrared region, ${\bf Y}_{lm}(\theta,\phi)$ is a spherical vector 
harmonic, $K_j(s)$ is expressed in terms of the confluent hypergeometric 
function
\begin{equation}
K_j(s) =s^j\,e^{-is}\,F\,(j-1,\, 2j+2,\, 2is),
\label{91}\end{equation}                                      
and $j$ runs through values which are positive roots of the equation 
\begin{equation}
j(j+1)=l(l+1)+4,\qquad l=1,2,\ldots .
\label{92}\end{equation}                                       
We are looking for solutions in the class of functions with an appropriate 
behavior at spatial infinity and singular points of the background.
Every solution corresponding to a negative root of Eq. (\ref{92}) is more 
singular than the background (\ref{40}) at $r=0$, and should be excluded.
The solution ${\bf b^+}$ is obtained from Eq. (\ref{90}) by replacing $t$ by $-t$.

$K_j(s)$ is regular at $s=0$ while, in the limit $s\to\infty$, it has the 
asymptotic
\begin{equation}
K_j(\omega r)=\bigl[c_j\omega r+d_j+O\,((\omega r)^{-1})\,
\bigr]\exp(i\omega r)
\label{93}\end{equation}                                         
where $c_j$ and $d_j$ are certain known constants. 

Note that the simultaneous presence of ${\bf b}^+$ and ${\bf b}^-$ 
ensures a nonvanishing contribution to the energy-momentum tensor.
From the behavior of ${\bf b}^\pm$ at spatial infinity, Eq. (\ref{93}), it 
follows that the four-momentum $P_\mu$ is infrared divergent, and the 
semiclassical treatment of the cold phase turns our to be inconsistent.
Therefore, scenarios of hadronization with the presence of free quarks 
must be ruled out. 

Let us turn to the problem of stability of the solution (\ref{43}).
We restrict ourselves to the static case which though is hard justified now 
if one remembers the runaway problem.
Thus we consider only necessary (not sufficient) conditions of stability.

Equation (\ref{43}) is independent of $g$, hence $\pm\,(2i/g)$ must be replaced by 
$q$ in each foregoing relation.
Equation (\ref{88}) converts to the form
\begin{equation}
\bigl(\Box
\mp{2igq\over r}\,{\partial\over\partial t}-{g^2q^2\over r^2}
\,\bigr)\,{\bf b}^\pm =0,
\label{95}\end{equation}                                     
Equations (\ref{91})--(\ref{93}) are modified appropriately,
\begin{equation}
K_j(s) =s^j\,e^{-is}\,F(-igq +j +1,\, 2j+2,\, 2is),
\label{96}\end{equation}                                     
\begin{equation}
j(j+1)=l(l+1) -g^2q^2,\qquad l=1,2,\ldots 
\label{97}\end{equation}                                        
\begin{equation}
K_j(s) = O\,(s^{igq -1}\,e^{is}),\qquad s\to\infty.
\label{98}\end{equation}                                      

It is clear from Eq. (\ref{98}) that $q$ must be {\it real} for ${\bf b}^\pm$ to
decrease as $1/r$ at spatial infinity, similar to the background (\ref{43}). 
Let us compare their behaviors at $r =0$. 
From Eq. (\ref{96}) it follows that $K_j(s)$ is regular at $s =0$ 
if $j\ge 0$. 
Write the positive solution of Eq. (\ref{97}), 
\[j ={1\over 2}(\sqrt{(2l +1)^2 -4g^2q^2} -1),\] 
and take $l=1$, the minimal allowable value, then one finds that $j$ is 
positive for
\begin{equation} 
g^2q^2\leq 2.
\label{99}\end{equation}                                     
A similar result was obtained by Mandula \cite{mand}.
Thus the solution (\ref{40}) is stable whereas the solution (\ref{43}) would 
be so provided that $q$ is a real quantity less than ${\sqrt 2}/g$.

We next go to the two-quark case. 
Decompose $b_\mu$ into vectors of the color basis 
(\ref{61})--(\ref{66}),
\[b_\mu=\sum_{n =1}^3\,\bigl[\,b_\mu^n\,H_n+\sum_{k=1}^3\,(\,
b_\mu^{kn-}\,E^-_{nk}+b_\mu^{kn+}\,E^+_{kn}\,)\bigr].\] 

We restrict ourselves to the situation of static quarks.
We adopt the gauge condition  
\[{b_0 =0,}\]
which ensures the constancy of the color charges of both quarks.
Let us consider the background potential (\ref{67}). 
Now, by repeating what was done in the single-quark case, we find that 
${\bf b}^n$ satisfy Eqs. (\ref{86}) and (\ref{87}) while ${\bf b}^{23\pm}$ and 
${\bf b}^{13\pm}$ satisfy 
Eqs. (\ref{86}), (\ref{88}) and (\ref{89}) with $r$ playing the role 
of $\rho_1$ for ${\bf b}^{23\pm}$ and $\rho_2$ for 
${\bf b}^{13\pm}$. 
From this identification, one checks the stability of these components.
Note that ${\bf b}^{23\pm}$ and ${\bf b}^{13\pm}$ are 
associated with the position of corresponding quarks while 
${\bf b}^n$ does not relate to either quark specifically.

As for ${\bf b}^{12\pm}$, it obeys Eq. (\ref{86}) and  
\begin{equation}
\bigl[\Box 
\mp 4\,({1\over r_2}-{1\over r_1})\,{\partial\over\partial t}
+4\,({1\over r_2}-{1\over r_1})^2\,\bigr]\,{\bf b}^{12\pm}=0,
\label{100}\end{equation}                                   
where the operator $\Box$ acts on the variables $t$ and ${\bf x}$, and 
$r_I=\vert {\bf x}-{\bf z}_I\vert$.
One can see that ${\bf b}^{12\pm}$ fluctuates with respect to both 
quarks, hence quark binding is ensured by just this component of excitations.
It is essential to gain insight into the behavior of 
solutions of Eq. (\ref{100}) at spatial infinity.

If the quarks are separated by distance $d$, then, for 
$r_I\gg d$, Eq. (\ref{100}) is reduced to the wave equation, and 
its asymptotical solutions are either 
\begin{equation}
{\bf b}^{12\pm}\sim{\rm const} 
\label{101}\end{equation}                                  
or 
\begin{equation}
{\bf b}^{12\pm}\sim\sum_{k,l,m}\,j_l(kr)\,[\,{\bf c}_
{lm}^\pm(k)\,Y_{lm}(\theta,\phi)\,e^{-i k t}
+{\bf d}_{lm}^\pm(k)\,Y_{lm}(\theta,\phi)^*e^{i k t}\,]
\label{102}\end{equation}                                   
where $j_l(kr)$ are the spherical Bessel functions
\begin{equation}
j_l(kr)\sim\frac{1}{kr}\,\sin(kr-\frac{\pi l}{2}),\quad kr\gg l.
\label{103}\end{equation}                                    
The solution (\ref{102}) poses no infrared problem.
By contrast, the solution (\ref{101}) gives rise to the infrared 
divergence of $P_\mu$ and should therefore be considered as {\it redundant}.

The analysis of stability of the background (\ref{76}) generated by 
$K$-quark clusters is identical to that in the two-quark case, with 
${\bf b}^{I\,K+1}_\pm,\,\,I=1,\ldots,K$ playing the role of ${\bf b}^{23\pm}$ 
and ${\bf b}^{13\pm}$, while ${\bf b}^{12\pm}$ being represented by 
${\bf b}^{J\,L}_\pm,\,\,J,L=1,\ldots,K$.
The last field fluctuates with respect to the pair of quarks labeled by the 
numbers $J$ and $L$, which ensures their binding.

The existence of excitations ${\bf b}^{I\,K+1}_\pm,\,\,I=1,\ldots,K$ entails 
the infrared divergences of $P_\mu$ due to their asymptotical 
behavior, Eq. (\ref{93}), and the situation cannot be remedied by mere 
selecting scenarios of hadronization.  
How to exclude such excitations, is not yet understood. 
A possible direction in which one might search is studying a nonlinear 
problem of stability with the requirement that every excitation becomes  
purely gauge at spatial infinity.

\section{Semiclassical treatment} 
We have seen that the linearly rising term of $A_\mu$ produces no 
confining force.
This brings up a question: Is Wilson's criterion fulfilled?
From the trace relations 
\[{\rm tr}\,E^\pm_{mn}={\rm tr}\,H_n={\rm tr}\,(E^\pm_{mn}\,
E^\pm_{mn})={\rm tr}\,(H_n\,E^\pm_{mn})=0,\]
it follows that the loop operator
\[W(C) ={\rm tr}\,P\,\exp\,[ig\oint_C dz^\mu\,A_\mu(z)]\]
with the background $A_\mu$ develops the perimeter law for both phases.

Consider the effect of gluon excitations $b_\mu$ about $A_\mu$ in the 
cold phase. 
Substitution of $B_\mu=A_\mu+b_\mu$ to the YM action gives
\[S(B)\approx S(A)+{1\over 2}\int\!d^4x\,d^4y\,b^a_\mu(x)\,{\delta^2 
S(A)\over\delta A^a_\mu(x)\,\delta A^b_\nu(y)}\,b^b_\nu(y).\]
Although $S(A)$ is divergent it should be discarded, just as in 
electrodynamics.
Due to the gauge invariance, the differential operator 
$\delta^2 S/\delta A^a_\mu\,\delta A^b_\nu$ is irreversible, and a 
gauge-fixing term of the Lagrangian is called for.
(It would be reasonable for the present purposes to use some linear 
gauge-fixing condition to avoid complications associated with the 
Faddeev-Popov ghosts).
Thereafter a certain nonsingular Lagrangian of gluon excitations results,
\[{\cal L}=b_\mu^a(x)\,\Lambda^{\hskip2mm\mu\nu}_{ab}(A;\,
\partial)\,b_\nu^b(x),\]
where $\Lambda^{\hskip2mm\mu\nu}_{ab}(A;\,\partial)$ is a 
background-dependent reversible differential operator. 

We average $W(C)$ over either of two complex-conjugate backgrounds $A_\mu^+$ 
or $A_\mu^-$,  
\[\int {\cal D}\,b_\mu^c\,\exp\,[-\int d^4x\,b_\mu^a(x)\,
\Lambda^{\hskip2mm\mu\nu}_{ab}(A;\,\partial)\,b_\nu^b(x)
+i\oint_C dz^\mu\,b_\mu^b(z)\,].\]
This integral can be readily worked out to yield
\begin{equation}
\exp\,[-\oint_C dy^\mu \oint_C dz^\nu\,{\cal G}_{\mu\nu}^
{\hskip2mm ab}(y,\,z),\,]
\label{106}\end{equation}                                 
where ${\cal G}_{\mu\nu}^{\hskip2mm ab}(y,\,z)$ is the gluon 
propagator obeying the equation
\[\Lambda^{\hskip2mm\lambda\mu}_{ab}(A;\,\partial)(y)\,{\cal 
G}_{\mu\nu}^{\hskip2mm bc}(y,\,z)=-\delta^c_a\,\delta^
\lambda_\nu\,\delta^4(z).\]
The area law for Eq. (\ref{106}) would be the case if 
${\cal G}_{\mu\nu}^{\hskip2mm ab}(y,\,z)$ tends to a constant as 
$(y-z)^2\to-\infty$. 
Since the behavior of the propagator at spacelike infinity is the same as 
that of excitations $b_\mu^a$ which obey the corresponding homogeneous 
equations, the responsibility for the area law rests with the excitations 
${\bf b}^{J\,L}_\pm$ approaching asymptotically a constant as $r\to\infty$, 
the redundant solution of Eq. (\ref{100}). 
Thus Wilson's criterion is fulfilled, though the area law cannot already be 
interpreted as an evidence of the classical attractive constant force between 
quarks composing a cluster.

Note also that the area law stems from the excitations described by the 
redundant solution rendering $P_\mu$ infrared divergent.
This resembles the situation with the energetical criterion valid for 
the prerenormalization stage when $P_\mu$ is still ultraviolet divergent.

We learnt from the exact YM solutions that every classical cluster 
has a certain nonzero color charge.
The situation reverses on the semiclassical level where the color neutralness 
of clusters is attained on the average of the gluon vacuum.

Given a $K$-quark cluster, one may define the gluon vacuum as a state with no 
excitation about the background generated by this cluster. 
This state is represented by a vector of Hilbert space $\Psi$ such that the 
expectation value of every color-invariant quantity coincides with its 
classical value. 
However, there are invariants which are finite and of different 
signs for the complex-conjugate potentials $A_\mu^+$ and $A_\mu^-$, as 
exemplified by $C_3={\rm tr}\,(F_{\lambda\mu}\,F^\mu_{\hskip1mm\nu}\,
F^{\nu\lambda})$ with the complex-conjugate solutions (\ref{76}).
Requiring the uniqueness of the vacuum expectation value of $C_3$, one has 
inevitably to assign $(\Psi, C_3\Psi)=0$. 
Let $\Psi_+$ and $\Psi_-$ be vectors of Hilbert space associated with $+$
and $-$ terms of Eq. (\ref{76}).
Being eigenvectors of the total color charge operator $\hat Q$ [defined by 
Eq. (\ref{13})], 
\[\hat Q\,\Psi_\pm = Q_{(\pm)}\,\Psi_\pm,\] 
they are mutually orthogonal, 
\[(\Psi_-,\Psi_+) =0.\]
If the gluon vacuum is defined as 
\begin{equation}
\Psi ={1\over 2}\,(\Psi_+ +\eta\,\Psi_-),
\label{104}\end{equation}                                        
where $\eta=\exp i\delta$ is an arbitrary phase factor, then one gets   
\[(\Psi,\,\hat Q\,\Psi)=0.\] 
Thus the condition of the color neutralness of $K$-quark clusters is met on 
the average of the gluon vacuum; the cluster finds itself partly at the state 
$\Psi_+$ with the color charge $Q_{(+)}=2i/g\,\sum_{I=1}^K\, H_I$ and partly 
at the state $\Psi_-$ with the color charge $Q_{(-)}=-2i/g\,\sum_{I=1}^K\,H_I$.

In the case of several clusters, the gluon vacuum is defined in a similar way.
For example, given two two-quark clusters, the construction 
\begin{equation}
\Psi={1\over 4}\,\bigl[\,\Psi(1_+,2_+)+\eta_{-+}\,\Psi(1_-,2_+)
+\eta_{+-}\,\Psi(1_+,2_-) +\eta_{--}\,\Psi(1_-,2_-)\,\bigr]
\label{105}\end{equation}                                        
ensures the color-neutralness of both clusters. 
Here, $\Psi(1_{\sigma 1},2_{\sigma 2})$ are vectors of Hilbert space 
associated with four solutions $A_\mu$ given by Eq. (\ref{77}), $\sigma_J$ 
the sign of the color charge of the $J$th quark, and $\eta_{\sigma 1 \sigma 2}$ 
are arbitrary phase factors.

Were $A_\mu^-$ to be convertible to $A_\mu^+$ by a gauge transformation, any 
superposition of $\Psi_+$ and $\Psi_-$ such as that of Eqs. (\ref{104}) or 
(\ref{105}) would be forbidden from realization as a physical state due to 
the availability of some superselection rule \cite{st}.
Only the potential generated by free quarks can be converted to the 
complex conjugate by a gauge transformation.
There is no such transformation for $A_\mu$ generated by bound quarks, hence
the superselection rule does not occur, and every cluster remains 
color neutral.
  
We regard every $K$-quark cluster on the equal footing since the existence of 
multiquark clusters was revealed experimentally \cite{bal}.
It is well known, however, that hadrons are much more stable than multiquark 
clusters. 
One may wonder what a plausible explanation of this fact may be.

We can envision consecutive constructions of the YMW systems with the color 
spaces SL$(N,C)$, SO$(N,C)$, and Sp$(N,C)$.
There is nothing to decide between these alternatives, hence all should 
persist and interfere.
Is there the largest color cell outside of which three pictures become 
quite different?
Such a sell does correspond to the three-quark case.
For $n>4$, there are no isomorphisms between members of the series sl$(n,C)$, 
so$(2n-1,C)$, sp$(n,C)$, and so$(2n,C)$.
The interference of distinct color backgrounds is responsible for the
splitting of energetical levels, which leads to the decay of clusters.
No interference occurs in the single-quark case because
sl$(2,C)\sim{\rm so}(3,C)\sim{\rm sp}(1,C)$.
In the two-quark and three-quark cases, two alternatives interfere, 
respectively, sl$(3,R)$ and so$(3,2)\sim{\rm sp}(2,R)$, and 
sp$(3,R)$ and sl$(4,R)\sim{\rm so}(3,3)$.
Thus clusters with two or three quarks are moderately stable.
The interference of three alternatives keeps multiquark clusters 
away from stability. 

The color-neutralness of hadrons in the Gauss law sense may well be 
compatible with the observability of some specific color multiplet structure 
which reveals itself by infinite-dimensional unitary multiplets of SL$(4,R)$.
Dothan, Gell-Mann, and Ne'eman \cite{dgn} suggested that unitary multiplets of
SL$(3,R)$ are related to the Regge trajectories of mesons.
This group is generated by the angular momentum operators $L_i$ and the 
quadrupole operators $T_{ij}$ with the commutation relations
\begin{equation} 
[L_i,\,L_j]=i\epsilon_{ijk}\,L_k,
\label{107}\end{equation}                                 
\begin{equation} 
[L_i,\,T_{jk}]=i\epsilon_{ijl}\,T_{lk}+i\epsilon_{ikl}\,
T_{jl},
\label{108}\end{equation}                                 
\begin{equation} 
[T_{ij},\,T_{kl}]=-i\,(\delta_{ik}\epsilon_{jlm}+
\delta_{il}\epsilon_{jkm}+\delta_{jl}\epsilon_{ikm})\,L_m.
\label{109}\end{equation}                                 
The algebra sl$(3,R)$ represents the minimal scheme capable to explain 
two features of Regge trajectories: The $\Delta J=2$ rule and the 
apparently infinite sequence of hadronic states.

It was found in \cite{dgn} that two infinite unitary representations 
belonging to the ladder series
\[D^{\rm ladd}_{SL(3,R)}(0;R):\quad\{J\}=\{0,2,4,\ldots\},\]
\[D^{ladd}_{SL(3,R)}(1;R):\quad\{J\}=\{1,3,5,\ldots\},\]
are associated with the $\pi$ and $\rho$ trajectories.
Besides, there exists \cite{og} a unique spinorial ladder 
representation related to the $N$ trajectory
\[D^{\rm ladd}_{SL(3,R)}({1\over 2};R):\quad\{J\}=\{{1\over 2},
{5\over 2},{9\over 2},\ldots\},\] 
while the spinorial representation starting with $J=\frac{3}{2}$ belongs to 
the discrete series \cite{sj},
\[D^{\rm disc}_{SL(3,R)}({3\over 2};R):\quad\{J\}=\{{3\over 2},
{5\over 2},{7^2\over 2},{9^2\over 2},{11^2\over 2}\ldots\}.\]

Thus the SL$(3,R)$ scheme of Dothan, Gell-Mann and Ne'eman, 
being usefully applied to the Regge trajectories of mesons,
turns out to be inadequate to account for those of baryons. 

Ne'eman and {\v S}ija{\v c}ki \cite{ne} assumed that matters 
can be improved by a simultaneous application of sl$(3,R)$ and so$(1,3)$. 
The commutation relations can  be closed by embedding two algebras in 
sl$(4,R)$, a relativistic generalization of sl$(3,R)$.

With adopting this SL$(4,R)$, one can classify the SU$(3)_f$ octet states
acording to the $D^{\rm disc}_{SL(4,R)}({1\over 2},0)\oplus D^{\rm disc}_{SL(4,R)}
(0,{1\over 2})$ representation while the symmetrized product of this 
reducible representation and the finite-dimensional SL$(4,R)$ representation 
$(\frac{1}{2},\frac{1}{2})$ is used for the decuplet states.
Although this scheme is quite restrictive, it is in a good agreement with 
known data of hadronic spectroscopy and predicts several new states \cite{ne}.
The present exact solutions show that the gauge symmetry of the background 
generated by clusters composed of two or three quarks is just SL$(4,R)$.
\footnote{The isomorphism sl$(4,C)\sim{\rm so}(6,C)$ renders selected this gauge 
symmetry. 
Indeed, if the conformal extension of Minkowski space $M^{\#}$ is to be mapped 
in a topologically nontrivial way into the color space, then the color space 
SL$(4,C)$ is favored over other images since it has a real form
SL$(4,R)\sim{\rm SO}(3,3)$ isomorphic to the conformal group of the 
pseudo-Euclidean space $E_{2,2}$.}

A basis of sl$(4,R)$ contains six antisymmetric elements $M_{\mu\nu}$ and nine 
symmetric elements $T_{\mu\nu}$, $\mu,\nu=1\ldots3$, which can be regrouped in 
the subsets: $L_i={1\over 2}\epsilon_{ijk}\,M_{jk}$, $K_i=M_{0i}$, $T_{ij}$, 
$N_i=T_{0i}$, $T_{00}$, satisfying the commutation relations 
(\ref{107})--(\ref{109}) together with 
\begin{equation} 
[K_i,\,K_j]=-i\epsilon_{ijk}\,K_k,\quad [N_i,\,N_j]=i
\epsilon_{ijk}\,N_k;
\label{110}\end{equation}                                 
\begin{equation} 
[L_i,\,K_j]=i\epsilon_{ijk}\,K_k,\quad [L_i,\,N_j]=i
\epsilon_{ijk}\,N_k,
\label{111}\end{equation}                                 
\begin{equation} 
[K_i,\,N_j]=-i\,(T_{ij} +\delta_{ij}\,T_{00});
\label{112}\end{equation}                                 
\begin{equation} 
[K_i,\,T_{jk}]=-i(\delta_{ij}\,N_k +\delta_{ik}\,N_j),
\label{113}\end{equation}                                 
\begin{equation} 
[N_i,\,T_{jk}]=-i(\delta_{ij}\,K_k +\delta_{ik}\,K_j),
\label{114}\end{equation}                                 
\begin{equation} 
[L_i,\,T_{00}]=[T_{ij},\,T_{00}]=0,\quad
[K_i,\,T_{00}]=-2iN_i,\quad [N_i,\,T_{00}]=-2iK_i.
\label{115}\end{equation}                                 

SL$(4,R)$ is thus split into several subgroups: SO$(4)={\rm SO}(3)\times{\rm
 SO}(3)$ 
the maximal compact sub\-group generated by $L_i$ and $N_i$, SO$(1,3)$ the 
Lorentz group generated by $L_i$ and $K_i$, SL$(3,R)$ the 
``three-volume''-preserving group generated by $L_i$ and $T_{ij}$, $R^+$ the 
noncompact Abelian group generated by $T_{00}$.

The subgroup SO$(4)$ is utilized as a 
basis with $J^P$ content of some $(j_1,\,j_2)$ representation:
\[J^P =(j_1+j_2)^P,\,(j_1+j_2-1)^{-P},\ldots,(|j_1-j_2|)^{\pm 
P}.\] 
The operator $T_{ij}$ shifts SO$(4)$-multiplets in $(j_1,j_2)$ by 
$\Delta j_{1,2}=2$ [see Eq. (\ref{108})], and the structure of Regge sequences 
is reproduced by such shifting.
A remarkable fact is that we have arrived at hadrons with different total 
angular momenta $J$, including the half-integer. 

However, in the present context, quarks have neither spin nor orbital 
momentum. 
In the limit $N\to\infty$, we deal with bound quarks moving along straight 
world lines.
Where did these higher angular momenta come from?
We suppose them to be built out of gluon degrees of freedom. 
Indeed, $M_{\mu\nu}$ and $T_{\mu\nu}$ are related to our color basis as follows
\begin{equation} 
M_{ij}\equiv-i(E^+_{ij}-E^-_{ij}),\quad K_j\equiv i(E^+_{0j}
-E^-_{0j}),
\label{118}\end{equation}                                 
\begin{equation}
T_{ij}\equiv-i(E^+_{ij}+E^-_{ij}),\quad N_j\equiv i(E^+_{0j}
+E^-_{0j}),
\label{119}\end{equation}                                 
\begin{equation}
T_{00}\equiv 2iH_0,\qquad T_{jj}\equiv -2iH_j. 
\label{120}\end{equation}                                 
It is conceivable that gluon excitations about the background with the 
SL$(4,R)$ color symmetry can manifest themselves as if their color 
degrees of freedom were converted into spin degrees of freedom described by
irreducible unitary representations of SO$(1,3)$.

A conversion of isospin into spin in gauge theories discovered 
in \cite{jare} seems to be of the direct relevance to our discussion.
This phenomenon has its origin in a combination of some singular gauge field 
of magnetic type, such as the magnetic field generated by a monopole, and an 
isospin-degenerate field which is the source of a Coulomb-like electric field.
The Li${\rm \acute e}$nard-Wiechert term of the background is analogous to 
the field of the magnetic monopole while the components of excitations 
${\bf b}^{IJ\pm},\,I,J=1,\ldots,3$ with the asymptotical behavior (\ref{101}) 
play the role of the long-range electric field. 
The rotation generators $M_{ij}$ related to $E^\pm_{ij}$ by Eq. (\ref{118}) 
occur in the term of the angular momentum independent of the radius of 
gyration.
 
Unfortunately, this similarity is not quite complete.
An external color field with an appropriate SL$(4,R)$ degeneracy generating 
a long-range counterpart of the initial background field can hardly be 
conceived in the present context. 
On the other hand, restricting the consideration to a pure YM system, one 
faces infrared divergences due to the excitations ${\bf b}^{I4\pm}$. 

\section{Acknowledgments}
In the course of this work I have benefited from numerous 
conversations with many people. 
I want to express my gratitude to them for encouragement and criticism.
With apologies for inadvertent oversights, a partial list 
includes B.A. Arbuzov, V.G. Bagrov, B.M. Barbashov, A.O. Barut, 
I.L. Buchbinder, G.V. Efimov, T. Goldman, M.V. Gorbatenko, B.L. Ioffe, 
G.S. Iroshnikov, A.N. Leznov, L.N. Lipatov, L. Lusanna, V.A. Novikov, 
L.B. Okun', V.A. Rubakov, G.K. Savvidi, I.V. Volobuev, and
V.Ch. Zhukovsky.
This work was supported in part by International Science 
and Technology Center under Project No. 208.



\begin{references}

\bibitem[*]{byline} Electronic address: 
${\rm boris\_0903@spd.vniief.ru}$

\bibitem{kosy} 
B.P. Kosyakov,  Theor. Math. Phys. {\bf 87}, 632 (1991).

\bibitem{kos} 
B.P. Kosyakov, Phys. Lett. B {\bf 312}, 471 (1993); Theor. Math. Phys. 
{\bf 99}, 409 (1994).

\bibitem{wong}
S.K. Wong,  Nuovo Cimento A {\bf 65}, 689 (1970).

\bibitem{frad} 
E.S. Fradkin and D.M. Gitman, Phys. Rev. D {\bf  44}, 3230 (1991);
U. Heinz, Phys. Lett. {\bf  144B}, 228 (1984).

\bibitem{emc} 
U. Stiegler, Phys. Rep. {\bf 277}, 1 (1996). 

\bibitem{ya} 
L.G. Yaffe, Rev. Mod. Phys. {\bf 54}, 407 (1982); S.R. Das, 
{\it ibid}. {\bf 59}, 235 (1987).

\bibitem{hooft} 
G. 't Hooft, Nucl. Phys. {\bf B72}, 461 (1974).

\bibitem{wit} 
E. Witten, Nucl. Phys. {\bf B160}, 57 (1979).

\bibitem{wils}
K.G. Wilson, Phys. Rev. D {\bf 10}, 2455 (1974).

\bibitem{ne} 
Y. Ne'eman and Dj. {\v S}ija{\v c}ki, Phys. Lett. {\bf  157B}, 
267 (1985); Phys. Rev. D {\bf 37},  3267 (1988); 
{\bf  47},  4133 (1993).

\bibitem{col} 
S. Coleman, J. Math. Phys. {\bf 7},  787  (1966).

\bibitem{jare}
R. Jackiw  and C. Rebbi, Phys. Rev. Lett. {\bf 36}, 1116  (1976);
P. Hasenfratz and G. 't Hooft, {\it ibid}. {\bf 36},  1119  (1976).

\bibitem{bbs} 
A. Balachandran, S. Borchardt, and A. Stern, Phys. Rev. D {\bf 17}, 3247 
(1978).

\bibitem{barr} 
A.O. Barut and R. R{\c a}czka, {\it Theory of Group 
Representations and Applications} (PWN -- Polish Scientific 
Publishers, Warszawa, 1977).

\bibitem{alf}
V. De Alfaro, S. Fubini, G. Furlan, and C. Rossetti, {\it Currents in Hadron 
Physics} (Noth-Holland, Amsterdam, 1973).

\bibitem{ccj} 
C.G. Callan, Jr., S. Coleman, and R. Jackiw, Ann. Phys. (N.Y.) {\bf 59}, 42 (1970).

\bibitem{r}
F. Rohrlich, {\it Classical Charged Particles} (Addison-Wesley, Reading, 
MA, 1965).

\bibitem{sy}
J.L. Synge, Ann. Matem. Pura Appl. {\bf 84}, 33 (1970).

\bibitem{k} 
B.P. Kosyakov, Sov. Phys. Usp. {\bf 35},  135  (1992).

\bibitem{jack}
R. Jackiw, L. Jacobs, and C. Rebbi, Phys. Rev. D {\bf 20}, 474 (1979); 
P. Sikivie and N. Weiss, {\it ibid}. {\bf  20}, 487 (1979). 

\bibitem{teit}
C. Teitelboim, Phys. Rev. D {\bf  1}, 1572 (1970). 

\bibitem{man} 
N.S. Manton, Nucl. Phys. {\bf B126},  525 (1977).

\bibitem{raj}
R. Rajaraman, {\it Solitons and Instantons} (Noth-Holland, Amsterdam, 1982).

\bibitem{jr}
R. Jackiw  and P. Rossi, Phys. Rev. D {\bf  21},  426 (1980).

\bibitem{kosya} 
B.P. Kosyakov,  Sov. J. Nucl. Phys. {\bf 55}, 1424 (1992). 

\bibitem{co} 
S. Coleman,  in 1981 International School of Subnuclear Physics 
"Ettore Majorana" (unpublished).

\bibitem{mand}
L.E. Mandula,  Phys. Rev. D {\bf 14}, 3497 (1976). 

\bibitem{st} 
R. Streater and A.S. Wightman, {\it PCT, Spin, Statistics And 
All That} (Benjamin, New York, 1964).

\bibitem{bal}  
A.M. Baldin, Sov. J. Part. Nucl. {\bf 8},  175 (1977).

\bibitem{dgn}
Y. Dothan, M. Gell-Mann, and Y. Ne'eman, Phys. Lett. {\bf 17}, 148 (1965) 

\bibitem{og} 
V.I. Žgievetskii and E. Sokachev, Theor. Math. Phys. {\bf 23}, 845 (1975).

\bibitem{sj} 
Dj. {\v S}ija{\v c}ki, J. Math. Phys. {\bf 16}, 298 (1975).

\end{references}
\end{document}